\documentclass[12pt,onecolumn,peerreview,draftcls]{IEEEtran}

\usepackage{url}
\usepackage{graphicx}
\usepackage{stmaryrd}
\usepackage{amssymb,amsmath}
\usepackage{multirow}
\usepackage{color,cite}
\usepackage{verbatim}

\newtheorem{lemma}{Lemma}

\newtheorem{proposition}{Proposition}

\newtheorem{theorem}{Theorem}
\newtheorem{observation}{Observation}

\ifodd 0
\newcommand{\rev}[1]{{\color{blue}#1}} 
\newcommand{\com}[1]{\textbf{\color{red} (COMMENT: #1)}} 
\newcommand{\comg}[1]{\textbf{\color{green} (COMMENT: #1)}}
\newcommand{\response}[1]{\textbf{\color{magenta} (RESPONSE: #1)}} 
\else
\newcommand{\rev}[1]{#1}
\newcommand{\com}[1]{}
\newcommand{\comg}[1]{}
\newcommand{\response}[1]{}
\fi

\begin{document}

\title{Attack Prevention for Collaborative Spectrum Sensing in Cognitive Radio Networks}

\author{Lingjie Duan$^{\ast}$, Alexander W. Min$^{\dagger}$, Jianwei Huang$^{\ast}$, Kang G. Shin$^{\dagger}$\\
$^{\ast}$Network Communications and Economics Lab, Dept. of Information Engineering, The Chinese University of Hong Kong, Hong Kong\\
$^{\dagger}$Real-Time Computing Laboratory, Dept. of EECS, The University of Michigan, Ann Arbor, MI 48109-2121\\
email:\{dlj008, jwhuang\}@ie.cuhk.edu.hk, \{alexmin, kgshin\}@eecs.umich.edu
\thanks{This work is supported by the General Research Funds (Project Number 412509) established under the University Grant Committee of the Hong Kong Special Administrative Region, China. Corresponding author: Jianwei Huang.}}

\maketitle

\begin{abstract}
Collaborative spectrum sensing can significantly improve the
detection performance of secondary unlicensed users (SUs). However,
the performance of collaborative sensing is vulnerable to sensing
data falsification attacks, where malicious SUs (attackers) submit
manipulated sensing reports to mislead the fusion center's decision
on spectrum occupancy. {Moreover, attackers may not follow the
fusion center's decision regarding their spectrum access.}
This paper considers a challenging attack scenario where multiple
{rational} attackers overhear \rev{all} honest SUs' sensing reports
and cooperatively maximize attackers' aggregate spectrum
utilization. We show that, without attack-prevention mechanisms,
honest SUs are unable to transmit over the licensed spectrum, and
they may further be penalized by the primary user for collisions due
to attackers' aggressive transmissions.
To prevent such attacks, we propose two novel attack-prevention
mechanisms with direct and indirect punishments. The key idea is to
identify collisions to the primary user that should not happen if
all SUs follow the fusion center's decision. Unlike prior work, the
proposed simple mechanisms do not require the fusion center to
identify and exclude attackers.
%
%
The direct punishment can effectively prevent all attackers from
behaving maliciously. The indirect punishment is easier to implement
and can prevent attacks when the attackers care enough about their
long-term reward.
%
\end{abstract}


\section{Introduction}
\label{sec:Intro} Cognitive radios enable secondary unlicensed users
(SUs) to opportunistically access licensed spectrum bands when they
are not being used by primary licensed users (PUs), and thus can
effectively improve spectrum utilization \cite{SHaykin2005}. As a
key technology for realizing opportunistic spectrum access while
protecting PU communications, spectrum sensing aims to detect the
presence or absence of a primary signal with high accuracy. To
provide sufficient protection, spectrum sensing must be able to
detect even a very weak primary signal, e.g., -20\,dB for a DTV
signal in the IEEE 802.22 WRANs \cite{Cordeiro2006}. To meet such
stringent requirement, researchers have proposed the use of
\emph{collaborative spectrum sensing} to improve detection
performance by exploiting sensor location diversity
\cite{Ghurumuruhan2, Letaief2009, CHLee,Min09a}. In collaborative
spectrum sensing, multiple sensors\footnote{We use the terms ``SU''
and ``sensor'' interchangeably throughout the paper.} sense the
spectrum individually and then report their sensing results to a
central node (i.e., fusion center) for a final decision on spectrum
occupancy.

Collaborative sensing, however, is vulnerable to critical attacks,
such as sensing data falsification attacks, while its detection is
difficult. In CRNs, sensors can be deployed in unattended and
hostile environments, and thus can be compromised by attackers.
Thus, compromised or malicious sensors can intentionally send
distorted sensing results to the fusion center in order to disrupt
the incumbent detection process \cite{RChen_Infocom,AWMIN,HLi}. Such
attacks can be easily launched due to the openness of the low-layer
protocols stacks of cognitive radio devices \cite{WXu}. However, it
is challenging for the fusion center to accurately validate the
integrity of sensing reports because of the two unique features in
spectrum sensing---unpredictability in wireless channel signal
propagations and lack of coordination between PUs and SUs. The
sensing data falsification attack will ultimately result in a waste
of spectrum opportunities (in the form of false alarms), and/or
excessive interference to the PU communications (in the form of
missed detections). Therefore, this poses a significant threat to
the implementation of cognitive radio technology, and thus calls for
efficient attack detection and prevention mechanisms.

In this paper, we consider an attack scenario in which multiple
attackers (i.e., compromised SUs/sensors) {\em cooperate} to
maximize their aggregate spectrum utilization in cognitive radio
networks (CRNs). Despite the serious threat posed by collaborated
attacks, attacker collaboration have not been fully considered in
CRNs. We focus on the particularly challenging attack scenario in
which attackers can overhear all honest SUs' sensing reports,
whereas the honest SUs are unaware of the existence of attackers.
This information asymmetry gives the attackers maximum capability to
launch attacks and achieve their goals. We design attack-prevention
mechanisms that safeguard collaborative sensing in such a
challenging attack scenario, which constitutes the main contribution
of this paper.

We consider two different attack scenarios: the ``attack-and-run''
scenario in which attackers only care about an immediate reward, and
the ``stay-with-attacks'' scenario in which attackers care about the
long-term reward. We first analyze the impact of attacks on honest
SUs in the absence of attack-prevention mechanisms. Then, we propose
two attack-prevention mechanisms: a {\em direct} punishment scheme
that can effectively prevent attacks in both scenarios mentioned
above, and an {\em indirect} punishment scheme that is easier to
implement and effectively prevents attacks in the
``stay-with-attacks'' scenario. The key idea of both mechanisms is
to discourage attackers from launching attacks by designing
efficient attack detection and punishment strategies.


The key results and organization of this paper are summarized as
follows.
\begin{itemize}
\item\emph{A spectrum-sharing model with collision penalty:}
In Sections~\ref{sec:NetworkModel} and \ref{sec:DataFusion}, we
introduce the concept of collision penalty, which requires the SUs
to compensate a PU for collision in utilizing the spectrum.
The collision penalty is designed to protect the PU's exclusive
spectrum usage and encourage the PU's opening of its licensed
spectrum to SUs.
\item\emph{Understanding cooperative attackers' optimal behaviors:}
In Section~\ref{sec:Game}, we theoretically show that in the absence
of attack-prevention mechanisms, attackers will utilize all spectrum
opportunities exclusively, whereas {honest SUs cannot transmit and
may even suffer from the collision penalty caused by attackers} (see
Table~\ref{tab:keyresult}).
\item\emph{Effective direct punishment:}
In Section \ref{sec:AttackPrevention1}, we design a direct
punishment mechanism that can detect attacks and punish the
attackers. {This requires an efficient way for the fusion center to
directly punish SUs.}
The proposed mechanism can {\em prevent all attacks} in both
``attack-and-run'' and ``stay-with-attacks'' scenarios (see
Table~\ref{tab:keyresult}). We further show that a single attacker
makes the network most vulnerable under this mechanism.
%
\item\emph{Effective indirect punishment:}
In Section~\ref{sec:AttackPrevention2}, we propose an indirect
attack-prevention mechanism that is easy to implement when direct
punishment is infeasible. The key idea is to terminate collaborative
sensing when an attack is detected.
The proposed mechanism can prevent all attacks if the attackers care
enough about their long-term reward (see Table~\ref{tab:keyresult}).
Unlike the direct punishment, the presence of a larger number of
attackers may make the network more vulnerable.
%
\end{itemize}

\begin{table*}
\renewcommand{\tabcolsep}{0.3cm}
\renewcommand{\arraystretch}{1.3}
\centering \caption{Key Results for different attack scenarios}
\begin{tabular}{|p{1.2in}||p{1.1in}|c|}
\hline \multicolumn{1}{|p{1.4in}||}{Attack Scenarios} &
\multicolumn{1}{|p{1.1in}|}{Attack-and-run} &
\multicolumn{1}{|c|}{Stay-with-attacks}\\\hline\hline
\multicolumn{1}{|p{1.4in}||}{No Punishment (Sec.\ref{sec:Game})} &
\multicolumn{2}{|c|}{Attacks happen and honest SUs {always} lose
transmission opportunities}\\\hline
\multicolumn{1}{|p{1.4in}||}{Direct Punishment
(Sec.\ref{sec:AttackPrevention1})} & \multicolumn{2}{|c|}{Completely
prevent attacks}\\\hline \multirow{2}*{Indirect Punishment
(Sec.\ref{sec:AttackPrevention2})} & \multirow{2}*{Cannot prevent
attacks} & \multirow{1}*{If attackers focus on long-term reward:
completely prevent attacks;}\\ & &If attackers focus on short-term
reward: partially prevent attacks.
\\\hline
\end{tabular}
\label{tab:keyresult}
\end{table*}

\subsection{Related Work}
\label{subsec:related}

%

There has been a growing interest in attack-resilient collaborative
spectrum sensing in CRNs (e.g.,
\cite{PKaligineedi,RChen_Infocom,AWMIN,HLi,OFatemieh}). Liu {\em et
al.} \cite{liu2009aldo} exploited the problem of detecting
unauthorized usage of a primary licensed spectrum. In this work, the
path-loss effect is studied to detect anomalous spectrum usage, and
a machine-learning technique is proposed to solve the general case.
Chen {\em et al.} \cite{RChen_Infocom} focused on a passive approach
with robust signal processing, and investigated robustness of
various data-fusion techniques against sensing-targeted attacks.
Kaligineedi {\em et al.} \cite{PKaligineedi} presented outlier
detection schemes to identify abnormal sensing reports.
Min {\em et al.} \cite{AWMIN} proposed a mechanisms for detecting
and filtering out abnormal sensing reports by exploiting
shadow-fading correlation in received primary signal strengths among
nearby SUs.
Fatemieh {\em et al.} {\cite{OFatemieh} used outlier measurements
inside each SU cell and collaboration among neighboring cells to
identify cells with a significant number of malicious nodes.} Li
{\em et al.} in \cite{HLi} detected possible abnormalities according
to SU sensing report histories.

Our work is different from existing approaches in three aspects.
First, we consider \emph{cooperation} among attackers, so the
attacks are much more challenging {to prevent}. {Second, unlike the
previous work which focused on sensing data falsification attacks,
we also consider the case where the attackers violate the fusion
center's decision regarding spectrum access.} Finally, our proposed
attack-prevention mechanisms can easily prevent attacks without
differentiating attackers from honest SUs.


%

\section{Preliminary}
\label{sec:NetworkModel}
\subsection{CRN Model and Assumptions}

We consider an infrastructure-based secondary CRN, which consists of
a single base station (or fusion center) and a set of SUs (or
sensors). The fusion center coordinates SUs' collaborative spectrum
sensing and their access to a licensed PU channel. We assume that
the fusion center is maintained by a trusted network administrator
and has high computation power. For collaborative spectrum sensing,
all SUs (i) measure the primary signal strength on the same target
channel, (ii) make local binary decisions on the presence or absence
of the primary signal, and (iii) report the binary decisions to the
fusion center \cite{Letaief2009,Peh:07}. Based on the reported
sensing results, the fusion center makes a global decision and
broadcasts this result to the SUs.

\begin{figure}[tt]
\centering
\includegraphics[width=0.5\textwidth]{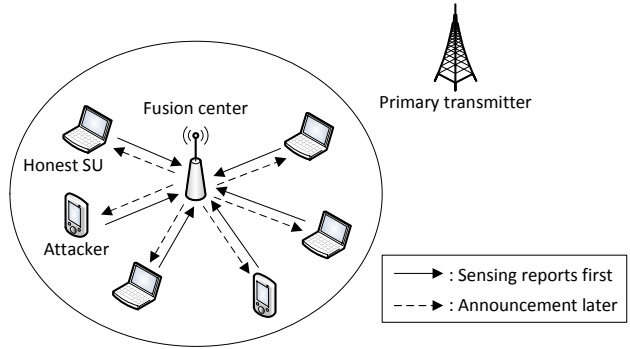}
\caption{{\bf An illustration of cooperative spectrum sensing in
cognitive radio networks}: The figure shows a secondary network with
$N=6$ SUs including $M=2$ malicious SUs (i.e., attackers). The SUs
periodically perform spectrum sensing and report the local (binary)
decisions to the fusion center (the solid arrows). The fusion center
makes a final decision and announces it to the SUs (the dotted
arrows).} \label{fig:network}
\end{figure}

There is a set of $\mathcal{N}=\{1,\ldots,N\}$ SUs in the network,
$M$ of which are attackers as shown in Fig.~\ref{fig:network}. We
assume that there is at least one honest SU in the network, i.e.,
\rev{$N-M \geq 1$}; otherwise, it would be infeasible to defeat
attacks. The honest SUs fairly share the licensed channel among
themselves when the channel is available to them (i.e., it is not
being used by the PU). The attackers (i.e., malicious or compromised
SUs), on the other hand, behave to maximize their own aggregate
reward (e.g., achievable throughput) by manipulating their sensing
reports so that the fusion center makes a wrong decision. In
particular, we focus on the case that attackers can overhear
\rev{all} honest SUs' sensing reports to the fusion center before
they collaboratively manipulating their sensing results. We assume
that attackers can communicate with each other (and thus know the
number of attackers), while the honest SUs only communicate with the
fusion center. \rev{The honest  SUs do not have to be strategic, and
they do not need to make decisions by considering other honest SUs
and attackers' decisions. In other words, the honest users do not
play a game with the attackers.}


To make the analysis tractable and obtain useful engineering
insights, we make the following assumptions throughout the paper.

\begin{description}
\item[\textbf{A1}.]
All SUs have the same detection performance in terms of primary
false alarm ($P_f$) and missed detection ($P_m$)
probabilities.\footnote{A false alarm occurs when an SU detects an
idle channel as busy, and a missed detection occurs when an SU
detects a busy channel as idle. The detection performance depends on
the SU's physical location (relative to the primary transmitter) and
fading environment. }
\item[\textbf{A2}.]
The PU's spectrum occupancy is the same for all SUs\footnote{This is
true when SUs stay relatively close compared to the PU's coverage
area.} and is independent across different time
slots.\footnote{\rev{This assumption is frequently used in the
literature (e.g., \cite{HLi,wang2009catchit,wang2009attack}), and is
reasonable when we try to approximate the case where PU's traffic
changes fast (e.g., wireless microphones) and the time slot is
relatively long. We may need to study the correlation between
spectrum occupancies when PU's traffic changes slowly over time
(e.g., TV transmitters). Analyzing the correlated case requires a
much more complicated Markov decision process (MDP) model than the
one that we used in Section~\ref{sec:AttackPrevention2}, and we
consider this as a future direction.}}
\item[\textbf{A3}.]
All SUs have the same transmission rate in utilizing the channel.
\end{description}

\rev{In Appendix~\ref{App:AssumpRelax}, we  relax both assumptions
\textbf{A1} and \textbf{A3} by studying SUs' heterogeneous detection
performances and heterogeneous transmission rates. We will focus on
the most challenging case of single attacker (as shown in
Theorem~\ref{prop:Cb_M} and Observation~\ref{ob:Cb_M_N} in
Section~\ref{sec:AttackPrevention1}), and show that the direct
punishment mechanism proposed in Section~\ref{sec:AttackPrevention1}
can still prevent all attacks. Similarly, the effectiveness of the
indirect punishment mechanism proposed in
Section~\ref{sec:AttackPrevention2} can also apply to the two
heterogeneous scenarios.}

Regarding the PU's temporal channel usage statistics, we denote
$P_I$ as the probability that the channel is \emph{actually} idle.
Thus, the channel is busy with the probability $1-P_I$. We assume
that SUs (including attackers and fusion center) know the
probability $P_I$ \rev{before collaborative spectrum sensing} as in
\cite{wang2009catchit,HLi,wang2009attack}. \rev{This is reasonable
if SUs and fusion center can collect PU's activity information from
PU side and calculate $P_I$ using various methods as in
\cite{Azzalini1981}. Such information collection is possible for SUs
by examining PU's published historical activity report or purchasing
the history report from PU directly. Actually, the precision of
$P_I$ does not affect SUs' decisions and our analytical results.
This is because attackers and the fusion center make decisions based
on their belief of $P_I$.}

\subsection{Spectrum Sensing and Opportunistic Access Model}

We assume a time-slotted model for opportunistic spectrum access, as
in Fig.~\ref{fig:timeslot}. Such time-slotted channel access
model has been widely assumed in the literature \cite{YCLiang,%
Zhao:07,Min10}, including the IEEE 802.22 standard draft
\cite{Cordeiro2006}. Each time slot consists of two phases:

\begin{figure}[tt]
\centering
\includegraphics[width=0.55\textwidth]{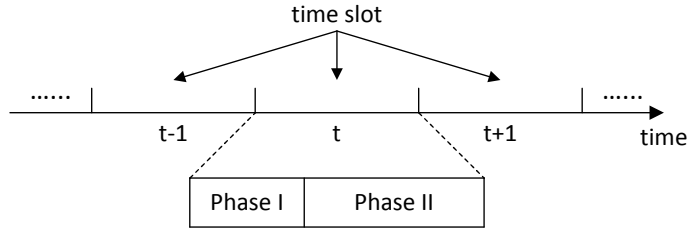}
\caption{{\bf The behaviors of SUs in each time slot}: Phase~I: SUs
sense and report, and then the fusion center fuses all SUs' reports
and announces the result; Phase~II: honest SUs transmit or wait
depending on the fusion center's decision. Attackers do not need to
follow the fusion center's decision.} \label{fig:timeslot}
\end{figure}

\begin{itemize}
\item Phase~I ({\it Collaborative Spectrum Sensing}): As shown in
Fig.~\ref{fig:network}, each SU performs sensing individually and
makes a local binary decision (i.e., 0/1) on channel occupancy: $1$
if it detects the PU's signal (i.e., busy), and $0$ otherwise (i.e.,
idle). All honest SUs truthfully report their sensing decisions to
the fusion center. The attackers, on the other hand, overhear the
sensing reports from the honest SUs before sending their own reports
(which may be different from their actual local sensing decisions)
to the fusion center. Based on the reports from all SUs (including
the attackers), the fusion center makes a global decision and
broadcasts it to all SUs in the network. We assume that the sensing
reports and announcements are communicated via a dedicated and
reliable control channel with no communication errors. \rev{Under
this one-hop network configuration, the attackers can overhear the
control channel and easily decode honest SUs' reports like the
fusion center.}\footnote{\rev{Note that end-to-end encryptions of
reports sent from SUs to the fusion center could be too complicated
and expensive to implement to prevent attackers' overhearing, as the
control channel often can only support very low date rate
transmissions.}} \rev{Also, even if we extend this one-hop
communication network to a multi-hop network, it is still possible
for attackers to overhear all honest SUs' reports as long as one
attacker is located near the fusion center.}

\item Phase~II ({\it Spectrum Sharing}): If the fusion center announces
the channel to be idle, then honest SUs will transmit in Phase~II.
If it announces the channel to be busy, then honest SUs will wait.
The attackers may transmit or wait in both cases. We assume that SUs
who transmit in Phase~II equally share the transmission time. More
advanced link scheduling and power control may improve the overall
network performance in Phase~II, but is not the focus of this paper.
Let us normalize the total transmission rate of the channel to
1.\footnote{If the total transmission rate of the channel is $r \,
(\neq 1)$, we can change $C_p$ and $C_b$ (defined later in the
paper) to $C_p/r$ and $C_b/r$ and all results will go through.}
{More specifically, ${X}$ SUs transmitting together leads to $1/{X}$
rate for each involved SU by using TDMA mode.}
\end{itemize}

{To summarize, the attackers can launch attacks in two different
ways: (i) in Phase~I by reporting falsified sensing results, and
(ii) in Phase~II by disobeying the fusion center's announcement.}


{
\subsection{Collision Penalty}\label{subsec:collision}


In order to increase social welfare, the government regulatory
bodies (e.g., FCC in the U.S. and Ofcom in the U.K.) are pushing new
spectrum-sharing schemes to allow the coexistence of PUs and SUs.
There are two main obstacles in persuading PUs to share their
licensed spectrum bands: (i) PUs' fear of interference or service
disruption caused by SUs, and (ii) lack of economic incentives to
PUs for spectrum sharing.
%
%
To achieve these goals while efficiently preventing attacks, we
adopt the notion of ``collision penalty'', similar in
\cite{huang2009optimal}, as an incentive mechanism to allow for an
efficient PU-SU coexistence. When a collision happens, we assume
that the PU will charge a collision penalty $C_p$ to {\em all} SUs
in the network. This collision penalty will compensate PUs for
potential performance loss due to
collisions.\footnote{\label{footnote:Cp}The penalty $C_p$ can be in
the form of monetary payments from SUs, or reduced transmission
opportunities of SUs, or cooperative transmission by SUs to improve
the PU's performance \cite{huang2009optimal,PricingBook}.} \rev{The
reasons why PU charges all SUs are as follows.
\begin{itemize}
\item \emph{Complexity consideration:} If the PU does not know each SU's transmission
characteristics (e.g., modulation and coding schemes), it is
impossible for him to check which subset of SUs cause collision.
Also, the attackers can secure their transmissions (e.g., via
MAC-layer encryptions) to avoid being detected and identified.
Moreover, it is highly complex and time-consuming for the PU to
identify which SUs cause usage collision. Such identification incurs
a detection delay and is thus not desirable
\cite{PKaligineedi,AWMIN,OFatemieh}.
\item \emph{Responsibility consideration:} In the cooperative spectrum sensing, each
SU contributes its sensing result to the final decision of the
fusion center, and each regular SU follows the final decision. If a
missed detection occurs, the PU should believe all SUs to be
responsible for their imperfect sensing. Even some missed detection
events are caused by attackers, it is impossible for the PU (without
sensing reports) to identify and punish attackers only.
\end{itemize}}
%

Based on the above discussion, we define the PU's expected utility
in one time slot as the sum of the PU's successful transmission rate
and collision penalty collected from $N$ SUs, i.e.,
\begin{equation}\label{eq:U_PU}
U_{PU}(C_p)=(1-\gamma(C_p))V(r_{PU})+\gamma(C_p) NC_p,
\end{equation}
where \rev{$\gamma(C_p)$ is the collision probability of the PU's
transmission due to SUs' aggressive access and is decreasing in
$C_p$, $r_{PU}$ is the PU's transmission rate, and $V(r_{PU})$ is
PU's utility of achieving rate $r_{PU}$}. A larger $C_p$ makes SUs
more conservative in spectrum access and leads to a lower
$\gamma(C_p)$. Hence, a larger $C_p$ achieves a high successful
transmission rate (in the first term in Eq.~(\ref{eq:U_PU})), but
may also lead to a low compensation from SUs (the second term in
Eq.~(\ref{eq:U_PU})).

%
%
%
%

}

\section{Decision Fusion Rule}
\label{sec:DataFusion} Of the various decision fusion rules for
collaborative sensing, we adopt the commonly used OR-rule. Ghasemi
and Sousa \cite{ghasemi2007opportunistic} showed that the OR-rule
performs better than other rules in many cases of practical
interest. Here we will discuss the OR-rule as a special case of the
general $n$-out-of-$N$ rule, and derive the conditions of $C_p$
under which the OR-rule is theoretically optimal. We elaborate the
decision fusion rule by focusing on the case in which all SUs are
honest.

At the end of Phase~I in each time slot (see
Fig.~\ref{fig:timeslot}), the fusion center collects a binary
sensing report $D_i \in \{0\,\textrm{(idle)},1\,\textrm{(busy)}\}$
from each SU $i \in \mathcal{N}$, and makes a decision using the
following $n$-out-of-$N$ rule \cite{Letaief2009}:
\begin{equation}
\label{eq:n_out_K}
\begin{cases}
\; \mathcal{H}_0 \; (\textrm{primary signal does not exist}):\
\text{if}\ \sum_{i\in\mathcal{N}}D_i<n. \\
\; \mathcal{H}_1 \; (\textrm{primary signal exists}):\ \text{if}\
\sum_{i\in\mathcal{N}}D_i\geq n
\end{cases}
\end{equation}

According to Eq.~\eqref{eq:n_out_K}, the fusion center infers the
channel to be busy $\mathcal{H}_1$ when at least $n$-out-of-$N$ SUs
report $1$ (busy); otherwise, it infers the channel to be idle
$\mathcal{H}_0$. The optimal selection of the threshold $n$ depends
on the system parameters and the reward functions of the SUs
\cite{Peh:07}. When $n=1$, we have the OR-rule.

We show that when both of the following conditions hold, the OR-rule
provides the highest reward for each SU within the family of
$n$-out-of-$N$ rules.
\begin{enumerate}
\item When all SUs report $0$ (i.e., $\sum_{i\in\mathcal{N}}D_i=0$),
each SU obtains a positive expected reward by sharing the spectrum
opportunity after taking into account the false alarm and missed
detection probabilities:
\begin{equation}\label{eq:OR_0}
Pr\left(\mathtt{idle}|\sum_{i\in\mathcal{N}}D_i=0\right)\frac{1}{N}-Pr\left(\mathtt{busy}|\sum_{i\in\mathcal{N}}D_i=0\right)C_p>0.
\end{equation}
The expected reward is the difference between the expected
transmission rate and the collision penalty. Here, $\mathtt{idle}$
and $\mathtt{busy}$ denote the actual state of the channel instead
of the fusion center's announcement (i.e., $\mathcal{H}_0$ or
$\mathcal{H}_1$).
\item When at least one SU reports $1$ (i.e., $\sum_{i\in\mathcal{N}}D_i\geq 1$),
every SU obtains a negative expected reward by sharing the spectrum
opportunity:
\begin{equation}\label{eq:OR_1}
Pr\left(\mathtt{idle}|\sum_{i\in\mathcal{N}}D_i\geq
1\right)\frac{1}{N}-Pr\left(\mathtt{busy}|\sum_{i\in\mathcal{N}}D_i\geq
1\right)C_p<0.
\end{equation}
\end{enumerate}

We can write these two conditions more compactly by defining the
following two notations:
\begin{align}\label{eq:H0_k}
P_{N,k}^I:&=Pr\left(\mathtt{idle}|\sum_{i\in\mathcal{N}}D_i=k\right)=\frac{P_I(1-P_f)^{N-k}P_f^k}{P_I(1-P_f)^{N-k}P_f^k+(1-P_I)(P_m)^{N-k}(1-P_m)^k},\\
P_{N,k}^B:&=Pr\left(\mathtt{busy}|\sum_{i\in\mathcal{N}}D_i=k\right)=1-P_{N,k}^I.
\label{eq:H1_k}
\end{align}

Notice that $P_{N,k}^I$ in Eq.~(\ref{eq:H0_k}) is decreasing in $k$,
and $P_{N,k}^B$ in Eq.~(\ref{eq:H1_k}) is increasing in $k$. Thus,
Eq.~(\ref{eq:OR_1}) is decreasing in $\sum_{i\in\mathcal{N}}D_i=k$.
This implies that with more SUs reporting $1$, SUs have less
incentive to transmit. We can summarize the range of $C_p$
satisfying both Eqs.~(\ref{eq:OR_0}) and (\ref{eq:OR_1}) as follows.

\vspace{0.2cm}
\begin{theorem}
At the fusion center, the OR-rule outperforms the other
$n$-out-of-$N$ rules ($n>1$) when the collision penalty $C_p$
satisfies the following condition.
\begin{align}
\label{eq:Cp_LBHB} \mathtt{Condition.I}:
\frac{P_I}{1-P_I}\left(\frac{1-P_f}{P_m}\right)^{N}
\frac{1}{N}\frac{P_mP_f}{(1-P_m)(1-P_f)}<C_p<\frac{P_I}{1-P_I}\left(\frac{1-P_f}{P_m}\right)^{N}\frac{1}{N}.
\end{align}
\end{theorem}

%

The lower-bound of $C_p$ in $\mathtt{Condition.I}$ discourages the
SUs from transmitting when at least one SU reports $1$ (busy). The
upper-bound of $C_p$ in $\mathtt{Condition.I}$ encourages SUs to
transmit when all $N$ SUs report $0$ (idle). In the rest of the
paper, we assume that $C_p$ always satisfies
{$\mathtt{Condition.I}$}.

\begin{figure}[tt]
\centering
\includegraphics[width=0.5\textwidth]{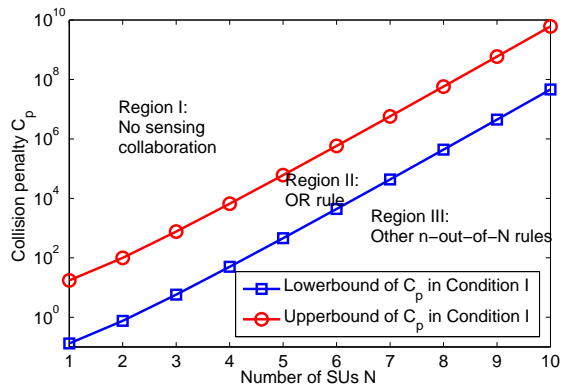}
\caption{$C_p$ range for the OR-rule's optimal application with
$(P_I,P_f,P_m)=(0.6,0.08,0.08)$.} \label{fig:C_pLBUB}
\end{figure}

\rev{In Region II of Figure~\ref{fig:C_pLBUB}, the OR-rule
outperforms the other $n$-out-of-$N$ rules with various number of
SUs and collision penalty $C_p$.\footnote{According to 802.22 WRAN
standard, $P_f$ and $P_m$ must be less than
10\%\cite{Cordeiro2006}.} As the number of SUs increases, the bounds
on $C_p$ increase. For a fixed $C_p$, more SUs lead to \rev{the
increase of false alarm probability and the decrease of missed
detection probability for the whole system}. Then the SUs tend to
strategically transmit more aggressively even when some SU(s)
reports $1$ (busy).  To prevent this and ensure the optimality of
the OR-rule, a higher value of $C_p$ is required. The other decision
fusion rules in Region III of Fig.~\ref{fig:C_pLBUB} is not the
focus of this paper. However, our analysis of the OR-rule can still
apply to Region III, \rev{since in later analysis we consider all
possible $C_p$ values and do not restrict our attention to
$\mathtt{Condition.I}$}.
}

\section{Attackers' Behaviors Without Punishment}
\label{sec:Game}

In this section, we analyze the behavior of cooperative attackers
when the system lacks attack-prevention mechanisms. The results in
this section will serve as a benchmark for the proposed
attack-prevention mechanisms in Sections~\ref{sec:AttackPrevention1}
and~\ref{sec:AttackPrevention2}.

We first define some useful notations.
\begin{itemize}
\item \emph{State set $\mathcal{S}$:}
A state $\boldsymbol{s}\in\mathcal{S}$ describes the local sensing
decisions of the honest SUs and attackers:
($\sum_{i\in\mathcal{N}\setminus\mathcal{M}}D_i,
\sum_{i\in\mathcal{M}}D_i$). The size of set $\mathcal{S}$ is
$(N-M+1)(M+1)$.\footnote{\rev{The value of $D_i$ can be either 0 or
1, thus $\sum_{i\in\mathcal{N}\setminus\mathcal{M}}D_i$ ranges from
0 to $N-M$ and $\sum_{i\in\mathcal{M}}D_i$ ranges from 0 to $M$.}}
The attackers know the exact state in a particular time slot by
overhearing the honest SUs' reports to the fusion center.
\item \emph{Attackers' action set $\mathcal{A}$:}
The action $\boldsymbol{a}_m$ of an attacker $m\in\mathcal{M}$ is a
tuple, (report to the fusion center {in Phase~I}, spectrum access
decision {in Phase~II}), which has 4 possibilities: (idle, wait),
(busy, wait), (idle, transmit), and (busy, transmit). Define
$\boldsymbol{a} = \{\boldsymbol{a}_m,\forall m\in\mathcal{M}\}$ as
the action vector of all attackers, and $\mathcal{A}$ includes all
possible $\boldsymbol{a}$.\footnote{Note that if all attackers have
the same action sets, the fusion center may find it easier to
identify them by checking their reports over time.}
\item \emph{Attackers' expected aggregate reward $R({\boldsymbol{a}},\boldsymbol{s})$:}
This reward depends on the state $\boldsymbol{s}$ and the attackers'
actions $\boldsymbol{a}$ in one time slot. \rev{It denotes the
difference between the attackers' aggregate transmission rate and
their expected payment to PU due to usage collision in one time
slot.}
\end{itemize}
For each state $\boldsymbol{s}$, the attackers choose
$\boldsymbol{a}$ to maximize the expected aggregate reward in a
single time slot, i.e.,
\begin{equation}
\label{eq:reward_nodelta} \max\limits_{\boldsymbol{a}\in\mathcal{A}}
R({\boldsymbol{a}},\boldsymbol{s}).
\end{equation}
We discuss the solution to Eq.~(\ref{eq:reward_nodelta}) in the
three following cases.

\subsection{All SUs sense the channel idle}
\label{subsec:1}

\begin{proposition}\label{thm:AD=0}
Given the state
$\boldsymbol{s}=\left(\sum_{i\in\mathcal{N}\setminus\mathcal{M}}D_i=0,
\sum_{i\in\mathcal{M}}D_i=0\right)$, the cooperative attackers'
optimal actions are: at least one attacker adopts the action (busy,
transmit) and the other attackers (if any) adopt the action (idle,
transmit). That is, at least one attacker will report the channel
busy in Phase~I and all attackers will transmit exclusively over the
channel in Phase~II. The fusion center will announce a wrong
decision $\mathcal{H}_1$ in this case. The attackers' expected
aggregate reward is:
\begin{equation}\label{eq:1-2}
R({\boldsymbol{a}},\boldsymbol{s}) = P_{N,0}^I-MP_{N,0}^BC_p>0,
\end{equation}
where the definitions of $P_{N,0}^I$ and $P_{N,0}^B$ are given in
Eqs.~(\ref{eq:H0_k}) and (\ref{eq:H1_k}), respectively. An honest SU
does not transmit, but may suffer from the collision penalty caused
by attackers and receives a negative expected reward
\begin{equation}\label{eq:1_honest}
R_{honestSU}(\boldsymbol{s})=-P_{N,0}^BC_p<0.
\end{equation}
\end{proposition}

\rev{ \emph{Proof.} Given the state $\boldsymbol{s} =
\left(\sum_{i\in\mathcal{N}\setminus\mathcal{M}}D_i=0,
\sum_{i\in\mathcal{M}}D_i=0\right)$ in a time slot, the attackers
may report truthfully and falsely in the Phase~I:
\begin{itemize}
\item If all attackers report $0$ (i.e., $\sum_{i\in\mathcal{M}}\tilde{D_i}
=\sum_{i\in\mathcal{M}}{D_i}=0$) in Phase I, then the announcement
at the fusion center is $\mathcal{H}_0$ and all honest SUs will
transmit. Consider ${M}_T$ ($0\leq {M}_T\leq M$) attackers choosing
to transmit rather than wait in Phase II. The attackers' expected
aggregate reward in this time slot is:
$$
R_{\boldsymbol{a}}(\boldsymbol{s})({M}_T)={M}_T\left(P_{N,0}^I\frac{1}{N-M+{M}_T}\right)-MP_{N,0}^IC_p,
$$
which is increasing in ${M}_T$. Thus all attackers will transmit,
i.e., ${M}_T=M$. Then the attackers' expected aggregate reward is
\begin{equation}\label{eq:1-1}
R_{\boldsymbol{a}}(\boldsymbol{s}) =
M\left(P_{N,0}^I\frac{1}{N}-P_{N,0}^BC_p\right)>0,
\end{equation}
due to {$\mathtt{Condition.I}$} in (\ref{eq:Cp_LBHB}).
\item If at least one attacker reports $1$ (i.e., $\sum_{i\in\mathcal{M}}\tilde{D_i}\geq 1$) in Phase I, then the announcement at the fusion
center is $\mathcal{H}_1$ and all honest SUs will not transmit.
\begin{itemize}
\item Consider ${M}_T$ ($1\leq {M}_T\leq M$)
attackers choosing to transmit. The attackers' expected aggregate
reward is given by (\ref{eq:1-2}), which does depend on $\tilde{M}$
and is larger than (\ref{eq:1-1}).
\item If all attackers wait, the attackers' expected aggregate reward
equals $0$, which is less than (\ref{eq:1-2}).
\end{itemize}
\end{itemize}
By comparing (\ref{eq:1-1}) and (\ref{eq:1-2}) with different
actions, we conclude that at least one attacker will report $1$ and
steal the opportunity from honest SUs to utilize the channel
exclusively. As a result, all honest SUs will not transmit but may
suffer the collision penalty as in (\ref{eq:1_honest}).
\hfill$\rule{2mm}{2mm}$ }

Proposition~\ref{thm:AD=0} shows that an attack always happens when
all SUs sense the channel idle.

\subsection{All honest SUs sense the channel idle, but some
attacker(s) senses the channel busy} \label{subsec:2} \rev{Here we
define the attackers' aggregate sensing result
$\sum_{i\in\mathcal{M}}D_i$ as $\bar{M}$.}
\begin{proposition}\label{thm:AD=1}
Given the state $\boldsymbol{s}=\left(\sum_{i\in\mathcal{N}\setminus
\mathcal{M}}D_i=0,\sum_{i\in\mathcal{M}}D_i=\bar{M}\geq 1\right)$,
the cooperative attackers' optimal actions are as follows.
\begin{itemize}
\item \emph{If $P_{N,\bar{M}}^I<MP_{N,\bar{M}}^BC_p$,} then at least one attacker
adopts the action (busy, wait) and the other attackers (if any)
adopt the action (idle, wait). This leads to a correct announcement
$\mathcal{H}_1$ (busy) at the fusion center. Since no one transmits,
the attackers and the honest SUs all get zero reward,
\begin{equation}
R({\boldsymbol{a}},\boldsymbol{s})=R_{honestSU}(\boldsymbol{s})=0.\end{equation}
\item \emph{If $P_{N,\bar{M}}^I\geq MP_{N,\bar{M}}^BC_p$,} then at least one
attacker adopts the action (busy, transmit) and the other attackers
(if any) adopt the action (idle, transmit). This leads to a correct
announcement $\mathcal{H}_1$ (busy) at the fusion center. Only
attackers will transmit exclusively in Phase~II, their expected
aggregate reward is:
\begin{equation}\label{eq:2-2}
R({\boldsymbol{a}},\boldsymbol{s}) =
P_{N,\bar{M}}^I-MP_{N,\bar{M}}^BC_p>0.
\end{equation}
An honest SU does not transmit in Phase II, but may suffer from the
collision penalty caused by attackers' transmissions and receives a
negative expected reward
\begin{equation}\label{eq:honest2}
R_{honestSU}(\boldsymbol{s}) = -P_{N,\bar{M}}^BC_p<0.
\end{equation}
\end{itemize}
\end{proposition}

\rev{ \emph{Proof.} Given the state
$\left(\sum_{i\in\mathcal{N}\setminus
\mathcal{M}}D_i=0,\sum_{i\in\mathcal{M}}D_i=\bar{M}\geq 1\right)$ in
one time slot, the attackers may report truthfully or falsely in
Phase I:
\begin{itemize}
\item If all attackers report $0$ (i.e., $\sum_{i\in\mathcal{M}}\tilde{D_i}
=\sum_{i\in\mathcal{M}}{D_i}=0$) in Phase I, then the announcement
at the fusion center is $\mathcal{H}_0$ and all honest SUs will
transmit. Similar to the proof in Subsection \ref{subsec:1}, it is
optimal for all attackers to transmit. Their expected aggregate
reward is:
\begin{equation}\label{eq:2-1}
R_{\boldsymbol{a}}(\boldsymbol{s}) =
M\left(P_{N,\bar{M}}^I\frac{1}{N}-P_{N,\bar{M}}^BC_p\right)<0,
\end{equation} due to {$\mathtt{Condition.I}$} in (\ref{eq:Cp_LBHB}).
\item If at least one attacker reports $1$ (i.e., $\sum_{i\in\mathcal{M}}\tilde{D_i}\geq 1$) in Phase I, then the announcement at the fusion
center is $\mathcal{H}_1$ and all honest SUs will not transmit.
\begin{itemize}
\item If at least one attacker transmits in Phase II, the
attackers' expected aggregate reward is given by (\ref{eq:2-2}).
Notice that (\ref{eq:2-2}) is negative only if the collision penalty
is high enough.
\item If all attackers wait in Phase II, the attackers' expected aggregate
reward equals $0$.
\end{itemize}
\end{itemize}

By comparing \eqref{eq:2-2} and $0$ with different actions, we
conclude that at least one attacker will  report $1$ to ensure that
the correct announcement is made at the fusion center. But the
attackers may transmit over the channel exclusively and the honest
SUs may suffer from the collision penalty caused by the attackers
with an expected reward in \eqref{eq:honest2}.
\hfill$\rule{2mm}{2mm}$ }

Proposition \ref{thm:AD=1} indicates that an attack only happens
when the benefit of exclusive transmission is large enough to
compensate the potential collision penalty for the attackers.

\subsection{Some honest SUs sense the channel busy}
\label{subsec:3}

\begin{proposition}\label{thm:AD=2}
Given the state $\boldsymbol{s}=\left(\sum_{i\in\mathcal{N}\setminus
\mathcal{M}}D_i=K\geq 1, \sum_{i\in\mathcal{M}}D_i=\bar{M}\geq
0\right)$,\footnote{\rev{Note that this state includes the case that
all honest SUs sense the channel busy and (some) attackers sense
idle.}} the announcement at the fusion center is always correct with
$\mathcal{H}_1$ (busy), and the attackers' optimal actions are as
follows.
\begin{itemize}
\item \emph{If $P_{N,K+\bar{M}}^I<MP_{N,K+\bar{M}}^BC_p$,} then each attacker
can either take the action (busy, wait) or (idle, wait). Since no
one transmits, the attackers and the honest SUs all get zero reward,
\begin{equation}
R({\boldsymbol{a}},\boldsymbol{s})=R_{honestSU}(\boldsymbol{s})=0.\end{equation}
\item \emph{If $P_{N,K+\bar{M}}^I\geq MP_{N,K+\bar{M}}^BC_p$,} then each
attacker can either take the action (busy, transmit) or (idle,
transmit). As only attackers will transmit in Phase~II, their
expected aggregate reward is:
\begin{equation}\label{eq:3-1}
R({\boldsymbol{a}},\boldsymbol{s}) =
P_{N,K+\bar{M}}^I-MP_{N,K+\bar{M}}^BC_p.
\end{equation}
An honest SU does not transmit in Phase~II, but may suffer from the
collision penalty caused by attackers' transmissions and receives a
negative expected reward
\begin{equation}\label{eq:honest3}
R_{honestSU}(\boldsymbol{s}) = -P_{N,K+\bar{M}}^BC_p<0.
\end{equation}
\end{itemize}
\end{proposition}

\rev{ \emph{Proof.} Given the state
$s=\left(\sum_{i\in\mathcal{N}\setminus \mathcal{M}}D_i=K\geq 1,
\sum_{i\in\mathcal{M}}D_i=\bar{M}\geq 0\right)$, no matter what
attackers report in Phase I, the announcement at the fusion center
is always $\mathcal{H}_1$, and the honest SUs will not transmit in
Phase II.
\begin{itemize}
\item If some attackers transmit in Phase II, the
attackers' expected aggregate reward is given by (\ref{eq:3-1}).
\item If all attackers wait in Phase II, the attackers' expected aggregate
reward equals $0$. Each honest SU's immediate expected reward also
equals $0$.
\end{itemize}

By comparing (\ref{eq:3-1}) to $0$, we conclude that the fusion
center always makes the correct announcement $\mathcal{H}_1$
regardless of the attackers' reports. However, the attackers may
transmit over the channel exclusively in Phase~II, and the honest
SUs may suffer from the collision penalty caused by the attackers
with expected reward in (\ref{eq:honest3}). \hfill$\rule{2mm}{2mm}$
}

\begin{table*}
\renewcommand{\tabcolsep}{0.3cm}
\renewcommand{\arraystretch}{1.3}
\centering \caption{Attackers' optimal behaviors and honest SUs'
behaviors}
\begin{tabular}{|p{0.92in}||c|p{0.65in}|}
\hline \multicolumn{1}{|p{0.92in}||}{Sensing Decisions} &
\multicolumn{1}{|c|}{Attackers' optimal behaviors} &
\multicolumn{1}{|p{0.65in}|}{Honest SUs' behaviors}\\\hline\hline
\multicolumn{1}{|p{0.92in}||}{$\sum_{i\in\mathcal{N}}D_i=0$} &
\multicolumn{1}{|c|}{Attack by reporting falsely and transmitting
exclusively} & \multicolumn{1}{|p{0.65in}|}{Wait} \\\hline
\multirow{2}*{$\sum_{i\in\mathcal{N}}D_i=K\geq1$} &
\multicolumn{1}{|c|}{If $P_{N,K}^I<MP_{N,K}^BC_p$, do not attack} &
\multirow{2}*{Wait}\\ & \multicolumn{1}{|l|}{If $P_{N,K}^I\geq
MP_{N,K}^BC_p$, attack by reporting truthfully and transmitting
exclusively}&
\\\hline
\end{tabular}
\label{tab:Attack}
\end{table*}


%

We summarize the results in
Propositions~\ref{thm:AD=0}-\ref{thm:AD=2} as in Table
\ref{tab:Attack}. Without any attack-prevention mechanism, the
attackers will utilize the spectrum opportunities exclusively,
whereas the honest SUs will never transmit regardless of their
sensing decisions. What is worse, the honest SUs may suffer from the
collision penalty caused by the attackers.


\vspace{0.2cm} \rev{Note that our current analytical results focus
on one time slot, where the attackers want to maximize their
expected aggregate reward in the current time slot (i.e., the
``attack-and-run'' scenario). Since attackers' behaviors are
independent over time slots, the above analytical results also hold
for the ``stay-with-attacks'' scenario.}

\rev{Given many possible attack scenarios in Section~\ref{sec:Game},
it is hard to identify attackers based on their report orders and
results in Phase~I. The reasons are as follows.
\begin{itemize}
\item First, different SUs may have different sensing times to guarantee
certain precision of channel detection, and thus it is not possible
to force everyone to report at the same time. This means that there
is always a last reporter. If all SUs are honest, then the last
reporter is not an attacker. Unless the fusion center is sure that
there exists at least one attacker, it is hard to tell that the last
reporting SU is an attacker.

\item Second, even the fusion center is aware of attacker(s), it is
still difficult to punish attackers effectively since the attackers
(aware of such identification) can strategically change to report
not the last.
\begin{itemize}
\item When the attackers overhear some honest SU(s) reporting 1
(busy) at the beginning, they can report immediately after their
sensing and do not need to wait for the last honest SU's report. In
this case, the fusion center's decision is correct ($\mathcal{H}_1$)
no matter attackers' manipulate their reports or not. But the
attackers can still attack (i.e., violating the fusion center's
decision and transmit) as shown in Case C in Section IV. In this
case, the attackers still need to overhear all honest SUs' reports.
\item When the attackers overhear many honest
SUs' reporting 0 (idle), they may not wait for the last honest SU's
report and can still manipulate their reports. In this case, such
identification still hurts honest SU(s) and the attackers still
perform attacks although they lose a little bit of information.
\end{itemize}
\end{itemize}

It should also be noted that it is possible for the fusion center to
monitor the control channel to check who are the attackers by
exchanging their sensing results secretly in Phase I. It is also
possible that the fusion center can monitor the PU's licensed later
to see who disobey its decision to transmit exclusively. But the
above attack identifications require the fusion center to know at
least all SUs' coding and modulation schemes. Even the fusion center
has such information, the attackers can still change their coding
and modulation schemes (e.g., as some honest SUs), or secure their
communication to exchange sensing results in Phase I and their
transmission in Phase II, e.g., via MAC-layer encryptions, to avoid
being identified by the fusion center.

The above discussions illustrate why we are interested in designing
attack-prevention mechanisms without attack identification. }

\section{Attack-Prevention Mechanism: A Direct Punishment}
\label{sec:AttackPrevention1}

In this section, we consider the case in which the fusion center can
directly charge a punishment to the SUs when attacks are identified.
We focus on the ``attack-and-run" scenario in a single time slot.
The analysis also applies to the ``stay-with-attacks" scenario as in
Section~\ref{sec:Game}. With the proper choice of punishment, the
proposed mechanism ensures that no attack will happen and no one
will be punished.

Let us denote the direct punishment as $C_b$, which is different
from the collision penalty $C_p$ introduced in
Section~\ref{subsec:collision}. The fusion center will only charge
the punishment to all SUs when the PU detects an attack. Let us
consider the following two scenarios:
\begin{itemize}\item \rev{When the announcement at the fusion center is $\mathcal{H}_1$ (busy)
in Phase~I and a collision happens in Phase II, the fusion center
knows that an attack happens (as honest SUs will not transmit in
Phase~II). In this case, all SUs are charged a direct punishment
$C_b$ by the fusion center (in addition to the collision penalty
$C_p$ charged by the PU).}\footnote{The way {for the fusion center}
to realize the punishment $C_b$ is similar to the way to realize the
collision penalty $C_p$. See footnote \ref{footnote:Cp} for
details.}
\end{itemize}
Note that when the announcement at the fusion center is
$\mathcal{H}_0$ (idle) in Phase~I, no direct punishment will be
triggered even if there is a collision in Phase~II. {This is because
attackers will not share the spectrum access opportunity with honest
SUs as in Proposition~\ref{thm:AD=0}, and such collision can only
the result of the missed detections of spectrum sensing.}


The effectiveness of the attack-prevention mechanism depends on the
choice of the punishment $C_b$. Theorem \ref{prop:Cb_M} shows that a
large enough $C_b$ can prevent all possible attacks.

\vspace{0.2cm}
\begin{theorem}
\label{prop:Cb_M} For $M$ attackers in the network, there exists a
threshold $C_b^{th}(M)$, i.e.,
\begin{equation}
\label{eq:Cb_2}
C_b^{th}(M)=\frac{P_I}{1-P_I}\left(\frac{1-P_f}{P_m}\right)^{N}
\max\bigg(\frac{P_fP_m}{(1-P_f)(1-P_m)}\frac{1}{M}-{C_p},\left(\frac{1}{M}-\frac{1}{N}\right)\bigg),
\ \ \forall M\geq 1,
\end{equation}
such that any value $C_b>C_b^{th}(M)$ can prevent all attack
scenarios described in Section~\ref{sec:Game}.
\end{theorem}

\begin{figure}[tt]
\centering
\includegraphics[width=0.5\textwidth]{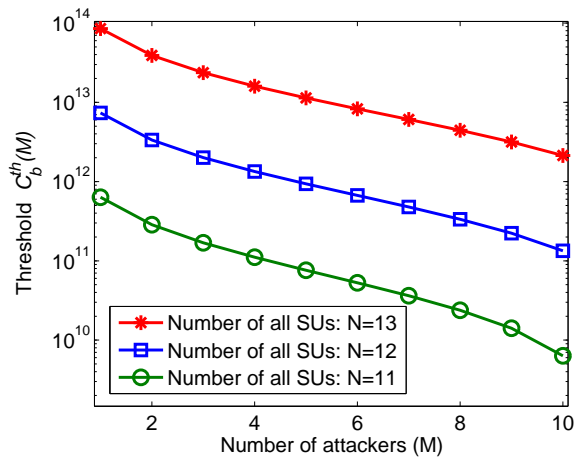}
\caption{Direct punishment threshold $C_b^{th}(M)$ for different $M$
and $N$ cases with ($P_I,P_f,P_m,C_p$)=($0.6,0.08,0.08,6e+10$).}
\label{fig:Cb_M_A}
\end{figure}

\begin{figure}[tt]
\centering
\includegraphics[width=0.5\textwidth]{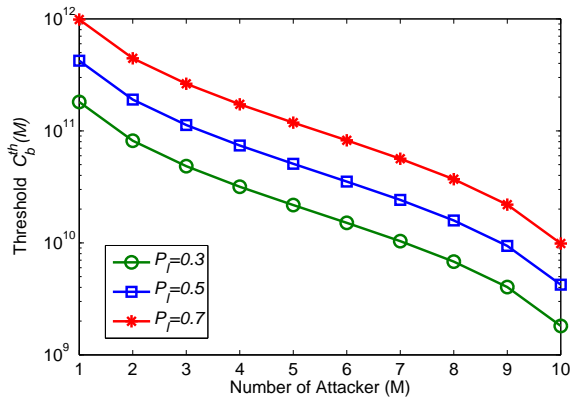}
\caption{Direct punishment threshold $C_b^{th}(M)$ for different $M$
and $P_I$ cases with ($P_f,P_m,C_p,N$)=($0.08,0.08,6e+10,11$).}
\label{fig:Cb_M_PI}
\end{figure}

\vspace{0.2cm}The proof of Theorem \ref{prop:Cb_M} is given in
Appendix~\ref{subsec:proof_theorem3}.
Next, we examine how the numbers of honest SUs and attackers affect
the threshold $C_b^{th}(M)$.

\vspace{0.2cm}
\begin{observation}\label{ob:Cb_M_N}
$C_b^{th}(M)$ is decreasing in the number of attackers $M$ and
increasing in the number of honest SUs $N-M$. If the fusion center
does not know the number of attackers, it should set the threshold
to be $C_b^{th}(1)=\max_{M\geq 1}C_b^{th}(M)$ to prevent all
attacks.
\end{observation}

\vspace{0.2cm} Figure~\ref{fig:Cb_M_A} shows the value of threshold
$C_b^{th}(M)$ as a function of $M$ for different values of
$N$.\footnote{\label{footnote:PfPm}\rev{Since $P_f$ and $P_m$ must
be less than $10\%$ in 802.22 WRAN standard draft, thus the
probability to trigger direct punishment is very small under this
choice of $P_f$ and $P_m$. As a result, high
$C_b^{th}(M)=C_b^{th}(M)/r$ value is determined in
Fig.~\ref{fig:Cb_M_A} to eliminate the attack benefit.}} When the
number of attackers increases, the total penalty to the group of
attackers also increases when an attack is confirmed (while the
total transmission rate does not change), which discourages the
attacks to happen.

Figure~\ref{fig:Cb_M_A} also shows that $C_b^{th}(M)$ increases with
the number of honest SUs $N-M$ for any fixed $M$. This is because
the more honest SUs' sensing reports are overheard by the attackers,
the more accurately the attackers can estimate the actual channel
state, and thus more likely the attackers will launch an attack. As
a result, a higher $C_b$ is required to prevent attackers from
manipulating their sensing reports. Thus, the single attacker
scenario (i.e., $M=1$) is the most challenging case for this
attack-prevention mechanism.



\vspace{0.2cm}
\begin{observation}\label{ob:Cb_rB_PI_Cp}
The threshold $C_b^{th}(1)$ is increasing in the idle probability
$P_I$ and non-increasing in the collision penalty $C_p$.
\end{observation}


\vspace{0.2cm} \rev{Figure~\ref{fig:Cb_M_PI} shows that the value of
threshold $C_b^{th}(M)$ is increasing in the idle probability
$P_I$.} A larger $P_I$ means a higher channel availability, and thus
encourages the attackers to launch an attack so that they can
exclusively utilize the channel more frequently. A larger $C_p$
discourages the attackers from accessing the channel due to the
possibility of paying a large collision penalty.
%


\section{Attack-Prevention Mechanism: An Indirect Punishment}
\label{sec:AttackPrevention2} The direct punishment scheme may be
difficult to enforce for certain types of networks due to practical
constraints, such as implementation overhead and complexity. {For
example, if the direct punishment is in the form of monetary
payments from SUs to the fusion center, the fusion center needs to
have reliable channels to collect and monitor such payments
\cite{shen2007optimal,PricingBook}.} In this section we propose an
indirect punishment scheme that can effectively prevent attacks in
the ``stay-with-attacks'' scenario as long as the attackers care
enough about future rewards.
The key idea is to terminate collaborative sensing once the fusion
center detects an attack, which forces the attackers to rely on
their own sensing results in the future. This prevents attackers
from overhearing honest SU sensing reports, and results in an
increase in missed detection probability for attackers.
Therefore, such indirect punishment will reduce the attackers'
incentives to attack.

The indirect punishment works as follows:
\begin{itemize}
\item \rev{When the fusion center announces $\mathcal{H}_1$ (busy) in Phase I
and a collision happens in Phase II, the indirect punishment is
triggered and there is no collaborative sensing in future time
slots.}\footnote{The fusion center can achieve this by broadcasting
to all SUs that there is no need to report local sensing decisions
in the future.}
\end{itemize}
Note that when the fusion center announces $\mathcal{H}_0$ (idle) in
Phase I, no indirect punishment will be triggered even if there is a
collision in Phase~II.

Similar to the direct punishment mechanism in
Section~\ref{sec:AttackPrevention1}, no indirect punishment will be
triggered if all SUs behave honestly. The effectiveness of the
indirect punishment depends on the attackers' performance when they
are isolated from the honest SUs.

In the rest of the section, we make the following assumption:
%
\begin{equation}\label{eq:A4}
\textbf{A4}:\ \ \ \ \ C_p>\frac{P_I}{1-P_I}\frac{1-P_f}{P_m}.
\end{equation}
\textbf{A4} is derived from $P_{1,0}^I-P_{1,0}^BC_p<0$, which
implies that a single SU will not transmit based on its own sensing
decision (since it can be quite unreliable after the collaborative
sensing breaks down) even without interference from the other SUs.
{\textbf{A4} is quite mild.} When the number of SUs is reasonable
(i.e., $N>7$), $\mathtt{Condition.I}$ in Eq.~(\ref{eq:Cp_LBHB})
directly guarantees the satisfaction of \textbf{A4} in
Eq.~(\ref{eq:A4}). Note that \textbf{A4} only applies to this
section.

%
To analyze the attackers' dynamic decisions in the long-term
``stay-with-attacks'' scenario, we formulate the problem as a Markov
decision process (MDP) \cite{Markov}. More specifically, we consider
an infinite horizon Markov decision process
$(\mathcal{S}',\mathcal{A}',P,R)$, where the group of cooperative
attackers is the only decision-maker (collectively) over time.
\begin{itemize}
\item \emph{State set $\mathcal{S}'$:} A state $\boldsymbol{s}
\in\mathcal{S}'$ describes the attackers' knowledge of honest SUs'
sensing decisions, their own sensing decisions, and whether the
indirect punishment is triggered:
($\sum_{i\in\mathcal{N}\setminus\mathcal{M}}\bar{D}_i$,
$\sum_{i\in\mathcal{M}}D_i$, $\mathtt{Punishment}$). When
$\mathtt{Punishment}=\mathtt{off}$,
$\sum_{i\in\mathcal{N}\setminus\mathcal{M}}\bar{D}_i=\sum_{i\in\mathcal{N}\setminus\mathcal{M}}{D}_i$.
When $\mathtt{Punishment}=\mathtt{on}$,
$\sum_{i\in\mathcal{N}\setminus\mathcal{M}}\bar{D}_i=\mathtt{Unknown}$
as the attackers do not know the honest SUs' sensing decisions. The
size of set $\mathcal{S}'$ is $[(N-M+1)(M+1)+(M+1)]$. The attackers
know the state during each time slot.

\item \emph{Attackers' action set $\mathcal{A}'$:} The action $\boldsymbol{a}_m$
of an attacker $m\in\mathcal{M}$ is a tuple: (report to the fusion
center, spectrum access decision). When the indirect punishment is
not triggered, there are four possible actions: (idle, wait), (busy,
wait), (idle, transmit), and (busy, transmit). When the indirect
punishment is triggered, an attacker's action can be ($N/A$,
transmit) or ($N/A$, wait), {where $N/A$ means that the attackers do
not report}. We define
$\boldsymbol{a}=\{\boldsymbol{a}_m,\forall m\in\mathcal{M}\}$ as the
action vector of all attackers and $\mathcal{A}^\prime$ contains all
feasible values of $\boldsymbol{a}$.

\item \emph{Transition probability $P({\boldsymbol{a}},\boldsymbol{s},\boldsymbol{s}')$:}
The transition probability that actions $\boldsymbol{a}$ in a state
$\boldsymbol{s}$ at time slot $t$ will lead to state
$\boldsymbol{s}'$ in time slot $t+1$ is
$P({\boldsymbol{a}},\boldsymbol{s},\boldsymbol{s}')=Pr(\boldsymbol{s}_{t+1}=\boldsymbol{s}'|\boldsymbol{s}_t=\boldsymbol{s},
\boldsymbol{a}_t=\boldsymbol{a})$. This depends on both state
$\boldsymbol{s}$ and actions $\boldsymbol{a}$, and is independent of
time $t$.
\item \emph{Attackers' expected aggregated reward $R({\boldsymbol{a}},\boldsymbol{s})$:}
The attackers' received reward after taking actions $\boldsymbol{a}$
in state $\boldsymbol{s}$ of a time slot.
\end{itemize}

Compared to the reward in the current time slot, the attackers may
value future rewards less. This can be captured by a discount factor
$\delta\in(0,1)$. We further define a stationary policy
$\boldsymbol{u}$ as a mapping between the set of states
$\mathcal{S}'$ to the action set $\mathcal{A}'$. In other words, a
policy defines what action to take in each possible state. The
attackers' objective is to choose a policy $\boldsymbol{u}$ from
policy set $\mathcal{U}$ to maximize the long-term expected
aggregate reward:
\begin{equation}\label{eq:reward_delta}
\max_{\boldsymbol{u}\in \mathcal{U}}\sum_{t=0}^\infty
\delta^tR({\boldsymbol{u}(\boldsymbol{s})},\boldsymbol{s}),
\end{equation}
Let us denote the attackers' optimal long-term expected aggregate
rewards by $LR^H$ and $LR^{DH}$ if they behave honestly and
dishonestly, respectively.


\rev{Since attackers' behaviors and rewards before and after the
indirect punishment are quite different, we need to study them
separately. Here we first consider the attackers' behaviors before
the punishment.} Let us consider the case where at least one SU
senses the channel busy, i.e., $\sum_{i\in\mathcal{N}}D_i=K\geq 1$.
The attackers' optimal behaviors can be classified into two cases:
\begin{itemize}
\item \emph{Non-aggressive Transmission:} The attackers will not attack
for any $K\geq 1$, which is true if
\begin{equation}
\mathtt{Case.NT:} P_{N,1}^I-MP_{N,1}^BC_p<0,
\end{equation}
\rev{where the attackers' exclusive transmission opportunity does
not compensate their collision penalty.}
\item \emph{Aggressive Transmission:} The attackers may attack even if $K\geq
1$, which is true if
\begin{equation}
\mathtt{Case.AT:} P_{N,1}^I-MP_{N,1}^BC_p\geq 0.
\end{equation}
\end{itemize}

In the rest of this section, we focus on $\mathtt{Case.NT}$ with
$M\geq 1$ attackers. The discussion for $\mathtt{Case.AT}$ with
$M\geq1$
 is given in Appendix~\ref{subsec:AA1}.


We analyze the conditions under which attacks can be completely
prevented via an indirect punishment. We first need to understand
the attackers' performance degradation once the indirect punishment
is triggered. Since the attackers are cooperative, they can always
exchange sensing information among themselves. \rev{Depending on
whether the attackers will transmit after the indirect punishment,
we have two cases:}
\begin{itemize}
\item \emph{Weak Cooperation:} The attackers will not transmit even when all attackers sense the channel
idle,
\begin{equation}
\label{eq:CaseA} \mathtt{Case.WC}:\ P_{M,0}^I-MP_{M,0}^BC_p\leq0.
\end{equation}
This means that the attackers feel that their own sensing results
are not reliable enough (with a high missed detection probability).
$\mathtt{Case.WC}$ also implies that the attackers will definitely
not transmit if one or more attackers sense the channel busy. {Due
to assumption \textbf{A4}, the reward in Eq.~(\ref{eq:CaseA}) is an
increasing function of the number of attackers $M$. Then we can also
write Eq.~(\ref{eq:CaseA}) as an upper bound of $M$, i.e.,
$\mathtt{Case.WC}$ corresponds to a small number of attackers $M$.}

\item \emph{Strong Cooperation:} The attackers will transmit when
all attackers sense the channel idle,
\begin{equation}
\label{eq:CaseIVB} \mathtt{Case.SC}:\ P_{M,0}^I-MP_{M,0}^BC_p>0.
\end{equation}
This means that the attackers feel that their own sensing results
(collectively) are accurate enough (with a low missed detection
probability) even taking the collision penalty $C_p$ into
consideration. {We can also write Eq.~(\ref{eq:CaseIVB}) as a lower
bound of $M$, i.e., $\mathtt{Case.SC}$ corresponds to a large number
of attackers $M$.}
\end{itemize}

Obviously, it is more challenging to prevent attacks in
$\mathtt{Case.SC}$ than $\mathtt{Case.WC}$. {However, we can show
that in $\mathtt{Case.SC}$ the attackers' expected aggregate reward
in one time slot with punishment triggered is always less than their
reward when they always behave honestly. }In other words, as long as
the attackers care enough about future reward (i.e., the discount
factor $\delta$ is high enough), we can still prevent attacks even
in $\mathtt{Case.SC}$ (and thus in $\mathtt{Case.WC}$ as well).
%


\vspace{0.2cm}
\begin{lemma}\label{lemma:longpunish1}
The attackers' optimal long-term expected aggregate rewards in
$\mathtt{Case.WC}$ and $\mathtt{Case.SC}$ are
\begin{equation}
LR_{WC}^H=LR_{SC}^H=Pr\left(\sum_{i\in\mathcal{N}}D_i=0\right)
\left(P_{N,0}^I\frac{1}{N}-P_{N,0}^BC_p\right)\frac{M}{1-\delta},
\label{eq:u_M_H} \end{equation}
\begin{equation}LR_{WC}^{DH}=
\frac{Pr(\sum_{i\in\mathcal{N}}D_i=0)(P_{N,0}^I-MP_{N,0}^BC_p)}
{1-\delta(1-Pr(\sum_{i\in\mathcal{N}}D_i=0)P_{N,0}^B)},
\label{eq:u_M_NH}
\end{equation}
and $LR_{SC}^{DH}$ in
\begin{align}
&LR_{SC}^{DH}=LR_{WC}^{DH}+\frac{\delta}{1-\delta}
Pr\Bigg(\sum_{i\in\mathcal{N}}D_i=0\Bigg)
\frac{P_{N,0}^BPr(\sum_{i\in\mathcal{M}}D_i=0)(P_{M,0}^I-MP_{M,0}^BC_p)}
{1-\delta(Pr(\sum_{i\in\mathcal{N}}D_i>0)+Pr(\sum_{i\in\mathcal{N}}D_i=0)P_{N,0}^I)}.
\label{eq:u_M_NH2} \end{align} Here the superscripts ``\emph{H}''
and ``\emph{DH}'' indicates honest and dishonest behaviors of
attackers, respectively.


\end{lemma}

\begin{table*}[t]
\begin{align}
&\delta^{th}_{SC}(M)=\Bigg(
1+\frac{(P_I(1-P_f)^N\frac{1}{N}-(1-P_I)(P_m)
^NC_p)-(P_I(1-P_f)^M\frac{1}{M}-(1-P_I)(P_m)^MC_p)}{\frac{1}{M}-\frac{1}{N}}
\frac{1-P_I}{P_I}\left(\frac{P_m}{1-P_f}\right)^N \Bigg)^{-1}.
\label{eq:delta2}
\end{align}
\hrule \vspace{-1cm}
\end{table*}

\vspace{0.2cm} The proof of Lemma \ref{lemma:longpunish1} is given
in Appendix~\ref{subsec:Lemma_proof_indirect}, where we can show
that $LR_{WC}^{DH}<LR_{WC}^{H}$ and $LR_{SC}^{DH}<LR_{SC}^{H}$ when
$\delta$ goes close to $1$. This leads to the following result.

\vspace{0.2cm}
\begin{theorem}\label{thm:IndirectDelta}
The indirect punishment can prevent all attack in
``stay-with-attacks'' scenario if the discount factor $\delta$
satisfies the following condition:
\begin{itemize}
\item \emph{Weak cooperation ($\mathtt{Case.WC}$):} for any $1\leq M<N$, we need $\delta>\delta_{WC}^{th}(M)$ where
\begin{equation}\label{eq:delta1}
\delta^{th}_{WC}(M)=\frac{1}{1+\frac{P_I(1-P_f)^N\frac{1}{N}-(1-P_I)(P_m)
^NC_p}{\frac{1}{M}-\frac{1}{N}}\frac{1-P_I}{P_I}\left(\frac{P_m}{1-P_f}\right)^N}.
\end{equation}
\item \emph{Strong cooperation ($\mathtt{Case.SC}$):} for any $1\leq
M<N$, we need $\delta>\delta_{SC}^{th}(M)$ where $\delta^{th}_{SC}$
is given in Eq.~(\ref{eq:delta2}).
\end{itemize}
{If the fusion center does not know the number of attackers, $M$, it
can choose $\delta>\max_{0<M<N}\delta^{th}_{WC}(M)$ and
$\delta>\max_{0<M<N}\delta^{th}_{SC}(M)$ for the two cases,
respectively.}
\end{theorem}

\vspace{0.2cm}

\vspace{0.2cm} Although it is not shown in
Theorem~\ref{thm:IndirectDelta}, we want to mention that the
indirect punishment can still partially prevent attacks even
$\delta$ is less than the discount factor threshold.
\rev{Intuitively, attackers do not want to trigger indirect
punishment and lose the opportunity to overhear honest SUs' sensing
results. Thus they will behave more conservatively compared to the
case with no indirect punishment. For example, if some SUs' sensing
results indicate a busy channel state, the attackers will not attack
to trigger the long-term punishment.}


\begin{figure}[tt]
\centering
\includegraphics[width=0.4\textwidth]{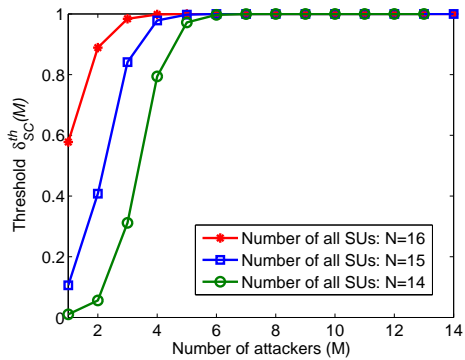}
\caption{Discount factor threshold $\delta_{SC}^{th}(M)$ with
$(P_I,P_f,P_m,C_p)=(0.6,0.08,0.08,3e+18).$} \label{fig:delta_M_N}
\end{figure}

We have the following interesting observations.


\vspace{0.2cm}
\begin{observation}\label{ob:longpunish_M}
(\emph{Impact of network size:}) Both $\delta^{th}_{WC}(M)$ in
$\mathtt{Case.WC}$ and $\delta^{th}_{SC}(M)$ in $\mathtt{Case.SC}$
are increasing in the number of the honest SUs $N-M$. Threshold
$\delta^{th}_{WC}(M)$ is decreasing in the number of the attackers
$M$, while $\delta^{th}_{SC}(M)$ is increasing in the number of the
attackers $M$.
\end{observation}

\vspace{0.2cm} Figure~\ref{fig:delta_M_N} plots
$\delta^{th}_{SC}(M)$ as a function of $N$ and $M$ in
$\mathtt{Case.SC}$. The corresponding result in $\mathtt{Case.WC}$
can also be obtained based on Eq.~(\ref{eq:delta1}).

With more honest SUs $N-M$, attackers have a less incentive to share
the spectrum with honest SUs in the long-term and a higher incentive
to attack and transmit exclusively. Thus a higher $\delta$ is needed
to prevent attacks.

A larger number of attackers $M$ has two effects: (a) a higher total
collision penalty (whenever a collision happens), and (b) attackers'
better estimation of channel condition (once the punishment is
triggered). It turns out that effect (a) dominates in
$\mathtt{Case.WC}$ and effect (b) dominates in $\mathtt{Case.SC}$,
which explains why the $\delta$ threshold decreases in $M$ in
$\mathtt{Case.WC}$ and increases in $M$ in $\mathtt{Case.SC}$.
In Fig.~\ref{fig:delta_M_N}, the most attack-vulnerable case happens
when almost all SUs are attackers ($M\rightarrow N$), in which case
$LR_{SC}^{DH}\rightarrow LR_{SC}^H$ and $\delta^{th}_{SC}(M)$ in
(\ref{eq:delta2}) is close to $1$.

%


\vspace{0.2cm}
\begin{observation}\label{ob:longpunish_r_Cp}
(\emph{Impact of collision penalty $C_p$:}) $\delta^{th}_{WC}(M)$ in
$\mathtt{Case.WC}$ is increasing in the collision penalty $C_p$,
while $\delta^{th}_{SC}(M)$ in $\mathtt{Case.SC}$ is decreasing in
$C_p$.
\end{observation}


\vspace{0.2cm} In $\mathtt{Case.WC}$, the collision penalty $C_p$
only affects the time slots before the punishment is triggered. A
higher $C_p$ means a smaller long-term expected reward as a
conservative honest SU (by comparing $LR_{WC}^{H}$ in
Eq.~(\ref{eq:u_M_H}) to $LR_{WC}^{DH}$ in Eq.~(\ref{eq:u_M_NH})),
and thus more incentives to attack. In $\mathtt{Case.SC}$, a larger
value of $C_p$ hurts the reward of attackers more after punishment
than before punishment. This is because the transmission probability
before punishment is $Pr(\sum_{i\in\mathcal{N}}D_i=0)$ (i.e., all
SUs sense idle), which is smaller than the transmission probability
after punishment $Pr(\sum_{i\in\mathcal{M}}D_i=0)$ (i.e., all
attackers sense idle). Thus a larger $C_p$ discourages the attacks
in $\mathtt{Case.SC}$.


\section{Conclusions and Future Work}
\label{sec:conclusion}

Collaborative spectrum sensing is vulnerable to sensing data
falsification attacks. In this paper, we focused on a challenging
attack scenario in which multiple cooperative attackers can overhear
the honest SU sensing reports, but the honest SUs are unaware of the
existence of attackers.
%
We first analyzed all possible attack scenarios without any
attack-prevention mechanisms. {In this case, we showed that honest
SUs will have no chance to transmit and may even suffer from the
collision penalty charged by the PU.} Then, we proposed two
attack-prevention mechanisms with direct and indirect punishments.
Both mechanisms do not require identification of the attackers. The
direct punishment can effectively prevent all attacks in both
``attack-and-run'' and ``stay-with-attacks,'' and the indirect
punishment can prevent all attacks in the long-run if the attackers
care enough about their future rewards.
%

There are several possible ways to extend the results in this
paper.\rev{
\begin{itemize}
\item First, we can consider the case where PU's traffic changes slowly
over time (e.g., TV transmitters), where we need to consider the
correlation between spectrum occupancies over different time slots.
In this case, we should to use a much more complicated MDP model
than the one in Section~\ref{sec:AttackPrevention2}.

\item Second, we can study the case that the fusion center knows all
SUs' transmission characteristics (e.g., modulation and coding
schemes) and can monitor attackers' sensing information
communication in Phase I and attackers' transmissions in Phase~II.
In this case, the fusion center may be able to identify attackers.
However, the attackers can change their modulation and coding
schemes \rev{(e.g., as some honest SUs)} or secure their
transmissions (via MAC-layer encryptions) to avoid being identified.

\item Third, we can study the denial-of-service attacks. Throughout this paper, we consider that
attackers are rational and are only interested in maximizing their
own rewards. For denial of service attacks, however, the attackers'
objective is to let honest SUs lose transmission opportunities or
break down the effectiveness of collaborative sensing.

\item Finally, we can consider imperfect control channel between SUs and the fusion
center (e.g., some SUs receive false announcement from the fusion
center). In that case, an indirect punishment can be triggered due
to channel communication errors instead of attacks. We need to
design the indirect punishment which will resume collaborative
sensing after a period of time (instead of an infinitely long
punishment).

\end{itemize}
}

\appendix

\subsection{Proof of Theorem
\ref{prop:Cb_M}}\label{subsec:proof_theorem3} By examining different
possible states, we can derive the attackers' optimal behaviors.
Then in response to the attackers' optimal behaviors, we find the
proper value of direct punishment $C_b$ to prevent all attacks.

\subsubsection{($\sum_{i\in\mathcal{N}\setminus\mathcal{M}}D_i=0$, $\sum_{i\in\mathcal{M}}D_i=0$)} When the sensing
results are all $0$, then the attackers may report truthfully or
falsely in the Phase I:
\begin{itemize}
\item If all attackers report $0$ in Phase~I, then the announcement at
the fusion center is $\mathcal{H}_0$ (idle), and all honest SUs will
transmit. It is easy to check that all attackers will transmit with
positive expected aggregate reward in (\ref{eq:1-1}).
\item If some attackers report $1$ in Phase~I, then the announcement at the fusion
center is $\mathcal{H}_1$ (busy) and all honest SUs will not
transmit.
\begin{itemize}
\item If some attackers with number $1\leq M_T\leq M$ choose to transmit, the
attackers' expected aggregate reward is:
\begin{equation}\label{eq:new_attackers1}
R_{\boldsymbol{a}}(\boldsymbol{s})=P_{N,0}^I-MP_{N,0}^B(C_p+C_b).
\end{equation}
which does not depend on $M_T$ and may not be larger than
(\ref{eq:1-1}).
\item If all attackers wait, the attackers' expected aggregate reward equals
$0$ and is less than (\ref{eq:1-1}).
\end{itemize}
\end{itemize}
To prevent attacks in this state, high value of $C_b$ should be set
to make (\ref{eq:1-1}) larger than (\ref{eq:new_attackers1}).

\subsubsection{($\sum_{i\in\mathcal{N}\setminus\mathcal{M}}D_i=0$, $\sum_{i\in\mathcal{M}}D_i=\bar{M}\geq 1$)}
The attackers may report truthfully or falsely in Phase~I:
\begin{itemize}
\item If all attackers report $0$ in Phase~I, then the announcement at
the fusion center is $\mathcal{H}_0$ (idle), and all honest SUs will
transmit. It is easy to check that all attackers will transmit and
their expected aggregate reward is given by (\ref{eq:2-1}) which is
negative.
\item If some attackers report $1$ in Phase~I, then the announcement at the fusion
center is $\mathcal{H}_1$ (busy), and all honest SUs will not
transmit.
\begin{itemize}
\item If some attackers with number $1\leq M_T\leq M$ choose to transmit, the
attackers' expected aggregate reward is:
\begin{equation}\label{eq:new_attackers2}
R_{\boldsymbol{a}}(\boldsymbol{s})=P_{N,\bar{M}}^I-MP_{N,\bar{M}}^B(C_p+C_b).
\end{equation}
which does not depend on $M_T$ and may be negative.
\item If all attackers wait, the attackers' expected aggregate reward equals
$0$ which is larger than (\ref{eq:2-1}).
\end{itemize}
\end{itemize}
To prevent attacks in this state, high value of $C_b$ should be set
to make (\ref{eq:new_attackers2}) smaller than $0$.

\subsubsection{($\sum_{i\in\mathcal{N}\setminus\mathcal{M}}D_i=K\geq
1$,
$\sum_{i\in\mathcal{M}}D_i=\bar{M}\in\{0,...,M\}$)}\label{subsec:3}

When at least one honest SU's sensing decision is $1$, then no
matter what attackers report in Phase~I, the fusion center always
makes correct announcement $\mathcal{H}_1$ (busy). All honest SUs
will not transmit in Phase~II.
\begin{itemize}
\item If some attackers with number $1\leq M_T\leq M$ choose to transmit in Phase~II, the
attackers' expected aggregate reward is:
\begin{equation}\label{eq:new_attackers3}R_{\boldsymbol{a}}(\boldsymbol{s})=P_{N,K+\bar{M}}^I-P_{N,K+\bar{M}}^B(C_p+C_b),\end{equation} which
does not depend on $M_T$ and can be positive or negative.
\item If all attackers wait in Phase~II, the attackers' expected aggregate
reward equals $0$.
\end{itemize}
To prevent all attacks in this state, high value of $C_b$ should be
set to make (\ref{eq:3-1}) smaller than $0$.

Then we can summarize the requirement of $C_b$ to prevent attacks in
all possible states in Theorem~\ref{prop:Cb_M}.

\rev{
\subsection{Relaxation of Assumptions
\textbf{A1} and \textbf{A3}}\label{App:AssumpRelax} Here we will
relax Assumption \textbf{A1} and consider the general case where SUs
have heterogeneous detection performances (i.e., different false
alarm probabilities $P_f$ and missed detection probabilities $P_m$)
and transmission rates. We are interested to know whether our
attack-prevention mechanisms (with some minor modification of system
parameters) can still apply, and how to change punishments to
attackers. Due to the page limit, we only examine the direct
punishment here. The effectiveness of the indirect punishment can be
shown similarly.

Observation~\ref{ob:Cb_M_N} in Section~\ref{sec:AttackPrevention1}
showed that the single attacker scenario ($M=1$) is the most
challenging attack scenario for the direct punishment mechanism.
Thus we will focus on the single attacker scenario to check the
effectiveness of this mechanism. We label the attacker as the $N$th
SU, and we denote its false alarm probability and missed detection
probability as $P_{f,A}$ and $P_{m,A}$, respectively. For the ease
of analysis, we still consider all honest SUs having the same $P_f$
and $P_m$.\footnote{If we consider different detection performances
for honest SUs, the analysis becomes more complicated without adding
more meaningful insights. \rev{Intuitively, as honest SUs' overall
detection performance becomes more precise, the attacker can predict
the channel state more precisely by overhearing honest SUs'
reports.}}

We denote the attacker's transmission rate as $r_A$, and an honest
SU $i<N$ has a different transmission rate $r_i$. When all SUs share
the same transmission opportunity using TDMA mode, the attacker
obtains a data rate of $r_A/N$.

First of all, we need to change the two notations in (\ref{eq:H0_k})
and (\ref{eq:H1_k}) as follows. When $0\leq k\leq N-1$, honest SUs
sense the channel busy $\left(\sum_{i=1}^{N-1}D_i=k\right)$ and the
attacker senses idle ($D_N=0$). The condition probability that the
channel is actually idle is
\begin{align}
P_{N,(k)+(0)}^I:&=Pr\left(\mathtt{idle}|\sum_{i=1}^{N-1}D_i=k,D_N=0\right)\nonumber\\
&=\frac{P_I(1-P_f)^{N-1-k}P_f^k(1-P_{f,A})}{P_I(1-P_f)^{N-1-k}P_f^k(1-P_{f,A})+(1-P_I)(1-P_m)^{k}P_m^{N-1-k}P_{m,A}}.
\end{align}
The conditional probability that the channel is actually busy is
\begin{equation}
P_{N,(k)+(0)}^B=1-P_{N,(k)+(0)}^I.
\end{equation}
When $0\leq k\leq N-1$, honest SUs sense the channel busy
$\left(\sum_{i=1}^{N-1}D_i=k\right)$ and the attacker also senses
busy ($D_N=1$). The conditional idle and busy probabilities are
respectively
\begin{align}
P_{N,(k)+(1)}^I:&=Pr\left(\mathtt{idle}|\sum_{i=1}^{N-1}D_i=k,D_N=1\right)\nonumber\\
&=\frac{P_I(1-P_f)^{N-1-k}P_f^kP_{f,A}}{P_I(1-P_f)^{N-1-k}P_f^kP_{f,A}+(1-P_I)(1-P_m)^{k}P_m^{N-1-k}(1-P_{m,A})},
\end{align}
and
\begin{equation}
P_{N,(k)+(1)}^B=1-P_{N,(k)+(1)}^I.
\end{equation}

With the help of above notations, we can similarly analyze how the
direct punishment works and how to determine the punishment as in
Section~\ref{sec:AttackPrevention1}.
\begin{theorem}\label{prop:Cb_M_new}
For the single attacker in the network, there exists a threshold
\begin{equation}
C_b^{th}=\frac{P_I}{1-P_I}\left(\frac{1-P_f}{P_m}\right)^{N-1}\frac{1-P_{f,A}}{P_{m,A}}\frac{N-1}{N}r_A,
\end{equation}
such that any value $C_b>C_b^{th}$ can prevent all attack scenarios
described in Section~\ref{sec:Game}.
\end{theorem}

\emph{Proof.} By examining different possible states as in
Section~\ref{sec:Game}, we can derive the attacker's optimal
behavior. Then in response to the attacker's optimal behavior, we
find the proper value of direct punishment $C_b$ to prevent all
attacks.

\subsubsection{State $\boldsymbol{s}=\left(\sum_{i=1}^{N-1}D_i=0, D_N=0\right)$} When the sensing
results are all $0$, then the attackers may report truthfully or
falsely in the Phase I:
\begin{itemize}
\item If the attacker reports $0$ in Phase~I, then the announcement at
the fusion center is $\mathcal{H}_0$ (idle), and all honest SUs will
transmit. It is easy to check that the attacker will also transmit
and receive a positive expected reward
\begin{equation}\label{eq:proof_RA_00}
R_A(\boldsymbol{s})=P_{N,(0)+(0)}^I\frac{r_A}{N}-P_{N,(0)+(0)}^BC_p>0,
\end{equation}
which is similar to (\ref{eq:OR_0}).
\item If the attacker reports $1$ in Phase~I, then the announcement at the fusion
center is $\mathcal{H}_1$ (busy) and all honest SUs will not
transmit.
\begin{itemize}
\item If the attacker chooses to transmit, its expected reward is
\begin{equation}\label{eq:proof_RA2_00}
R_A(\boldsymbol{s})=P_{N,(0)+(0)}^Ir_A-P_{N,(0)+(0)}^B(C_p+C_b),
\end{equation}
which may or may not be larger than (\ref{eq:proof_RA_00}).
\item If the attacker waits, its expected reward equals
$0$ and is less than (\ref{eq:proof_RA_00}).
\end{itemize}
\end{itemize}
To prevent attacks in this state, a high value of $C_b$ should be
set to make (\ref{eq:proof_RA_00}) larger than
(\ref{eq:proof_RA2_00}). In other words,
\begin{equation}\label{eq:proof_Cb1}
C_b>\frac{P_I(1-P_{f,A})}{(1-P_I)P_{m,A}}\left(\frac{1-P_f}{P_m}\right)^{N-1}\frac{N-1}{N}r_A,
\end{equation}
where we denote the term in the right-hand side as threshold
$C_b^{th1}$. It is easy to check that $C_b^{th1}$ is decreasing in
both $P_{f,A}$ and $P_{m,A}$, but is increasing in $P_I$, $N$, and
$r_A$.

\subsubsection{State $\boldsymbol{s}=\left(\sum_{i=1}^{N-1}D_i=0, D_N=1\right)$}
The attacker may report truthfully or falsely in Phase~I:
\begin{itemize}
\item If the attacker reports $0$ in Phase~I, then the announcement at
the fusion center is $\mathcal{H}_0$ (idle), and all honest SUs will
transmit. It is easy to check that the attacker will also transmit
and its expected reward is
\begin{equation}\label{eq:proof_RA_01}
R_A(\boldsymbol{s})=P_{N,(0)+(1)}^I\frac{1}{N}r_A-P_{N,(0)+(1)}^BC_b,
\end{equation}
which is negative as required by the optimality of OR-rule.
\item If the attacker reports $1$ in Phase~I, then the announcement at the fusion
center is $\mathcal{H}_1$ (busy), and all honest SUs will not
transmit.
\begin{itemize}
\item If the attacker chooses to transmit, its expected reward is
\begin{equation}\label{eq:proof_RA2_01}
R_A(\boldsymbol{s})=P_{N,(0)+(1)}^Ir_A-P_{N,(0)+(1)}^B(C_p+C_b),
\end{equation}
which may or may not be negative.
\item If the attacker waits, its expected reward equals
$0$ which is larger than (\ref{eq:proof_RA_01}).
\end{itemize}
\end{itemize}
To prevent attacks in this state, a high value of $C_b$ should be
set to make (\ref{eq:proof_RA2_01}) smaller than $0$. This gives to
\begin{equation}\label{eq:proof_Cb2}
C_b>\frac{P_IP_{f,A}}{(1-P_I)(1-P_{m,A})}\left(\frac{1-P_f}{P_m}\right)^{N-1}r_A-C_p,
\end{equation}
where we denote the term in the right-hand side as threshold
$C_b^{th2}$.

\subsubsection{State $\boldsymbol{s}=\left(\sum_{i=1}^{N-1}D_i=K\geq
1, D_N=\bar{M}\in\{0,1\}\right)$}

When at least one honest SU's sensing decision is $1$, then no
matter what the attacker reports in Phase~I, the fusion center
always makes correct announcement $\mathcal{H}_1$ (busy). All honest
SUs will not transmit in Phase~II.
\begin{itemize}
\item If the attacker chooses to transmit in Phase~II, its expected reward is
\begin{equation}\label{eq:proof_RA_KM}
R_A(\boldsymbol{s})=P_{N,(K)+(\bar{M})}^Ir_A-P_{N,(K)+(\bar{M})}^B(C_p+C_b),\end{equation}
which may be positive or negative.
\item If the attacker waits in Phase~II, its expected
reward equals $0$.
\end{itemize}
To prevent all attacks in this state, a high value of $C_b$ should
be set to make (\ref{eq:proof_RA_KM}) smaller than $0$. This gives
to
\begin{equation}
C_b>\frac{P_I}{1-P_I}\left(\frac{1-P_f}{P_m}\right)^{N-1}\left(\frac{P_fP_m}{(1-P_f)(1-P_m)}\right)^K\left(\frac{1-P_{f,A}}{P_{m,A}}\right)^{1-\bar{M}}\left(\frac{P_{f,A}}{1-P_{m,A}}\right)^{\bar{M}}r_A-C_p,
\end{equation}
which is maximized when $K=1$ and $\bar{M}=0$ due to $P_m<0.1$ and
$P_f<0.1$ as explained in footnote~\ref{footnote:PfPm}. To prevent
all attacks in this state, we should require
\begin{equation}
C_b>\frac{P_I}{1-P_I}\left(\frac{1-P_f}{P_m}\right)^{N-1}\left(\frac{P_fP_m}{(1-P_f)(1-P_m)}\right)\frac{1-P_{f,A}}{P_{m,A}}r_A-C_p,
\end{equation}
where the right-hand side is denoted as threshold $C_b^{th3}$.

To summarize, the requirement of $C_b$ to prevent attacks is
$C_b>\max\left(C_b^{th1},C_b^{th2},C_b^{th3}\right)$ in all possible
states. It is easy to check that $C_b^{th1}>C_b^{th2}$ and
$C_b^{th1}>C_b^{th3}$, and we can conclude the results in
Theorem~\ref{prop:Cb_M_new}.   \hfill$\rule{2mm}{2mm}$

\begin{observation}
$C_b^{th}$ is increasing in both idle probability $P_I$ and the
number of honest SUs $N-1$.
\end{observation}

A larger $P_I$ means a higher channel availability, and thus
encourages the attacker to launch an attack so that it can
exclusively utilize the channel more frequently. Also, as the honest
SU number $N-1$ increases, the more honest SUs' sensing reports are
overheard by the attacker. The attacker can estimate the actual
channel state more accurately, and it is more likely to launch an
attack.

\begin{observation}
Threshold $C_b^{th}$ is increasing in the attacker's rate $r_A$, and
it is decreasing in the attacker's false alarm probability $P_{f,A}$
and missed detection probability $P_{m,A}$.
\end{observation}

As the attacker's rate $r_A$ increases, it values the exclusive
transmission opportunity more. The attacker has a higher incentive
to attack, and a higher $C_b^{th}$ is required to prevent the
attack.

\begin{figure}[tt]
\centering
\includegraphics[width=0.5\textwidth]{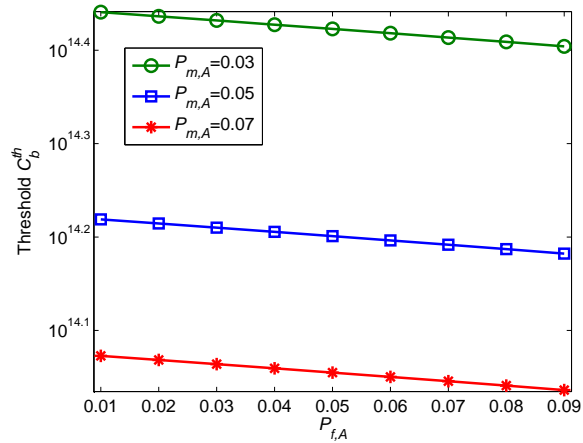}
\caption{Direct punishment threshold $C_b^{th}$ for different
$P_{f,A}$ and $P_{m,A}$ with ($P_f,P_m,N$)=($0.05,0.05,11$).}
\label{fig:Cb_PFA_PMA}
\end{figure}

Figure~\ref{fig:Cb_PFA_PMA} shows the threshold $C_b^{th}$ as a
function of the attacker's false alarm probability $P_{f,A}$ and
missed detection probability $P_{m,A}$. Intuitively, as $P_{f,A}$
increases, the attacker has higher probability to overlook the
channel access opportunity and it has less incentive to launch an
attack. As $P_{m,A}$ increases, the attacker has a higher
probability to trigger direct punishment and thus it is more
conservative to attack. In both cases, the fusion center can to
announce a lower $C_b$ to prevent all attacks. }

\rev{ \subsection{Attack Prevention in $Case.AT$ of Section
\ref{sec:AttackPrevention2}}\label{subsec:AA1}

Section~\ref{sec:AttackPrevention2} focuses on $\mathtt{Case.NT}$.
Here we discuss $\mathtt{Case.AT}$ with multiple attackers ($M\geq
1$). Table~\ref{tab:Attack} in Section~\ref{sec:Game} shows that
with no punishment, the attacker may still obtain a positive reward
by transmission even when at least one SU senses the channel busy in
$\mathtt{Case.AT}$. Next, we discuss the attacker's action
${\boldsymbol{u}}({\boldsymbol{s}})$ under any given state
$\boldsymbol{s}=\left(\sum_{i\in\mathcal{N}\setminus\mathcal{M}}\bar{D}_i,
\sum_{i\in\mathcal{M}}D_i, \mathtt{Punishment}\right)$.
\begin{itemize}
\item \emph{State $\boldsymbol{s}=\left(\sum_{i\in\mathcal{N}\setminus\mathcal{M}}D_i,\sum_{i\in\mathcal{M}}D_i=K, \mathtt{off}\right)$:} Before the indirect
punishment is triggered, the attackers may purposely change their
attack behaviors (comparing to the actions in Table~\ref{tab:Attack}
in Section~\ref{sec:Game}) in some states to deter punishment.
\begin{itemize}
\item \emph{$\boldsymbol{s}=\left(0, 0, \mathtt{off}\right)$:} at least one attacker still chooses the action (busy, transmit) as in
Table~\ref{tab:Attack}. In this state, the attackers obtain the
largest expected aggregate reward $P_{N,0}^I-MP_{N,0}^BC_p$ and
induces the smallest missed detection probability $P_{N,0}^B$ to
trigger punishment.
\item
\emph{$\boldsymbol{s}=(\sum_{i\in\mathcal{N}\setminus\mathcal{M}}D_i\geq0$,
$\sum_{i\in\mathcal{M}}D_i\geq0$, $\mathtt{off})$:} the attackers
may choose not to attack even if they can obtain a positive expected
aggregate reward in current time slot, which is different from no
punishment scenario in Table~\ref{tab:Attack}. This is because that
the attackers fear to trigger the long-term indirect punishment, and
they will attack only when missed detection probability to trigger
punishment is low with small $\sum_{i\in\mathcal{N}}D_i$.
\end{itemize}
\item \emph{State $\boldsymbol{s}=\left(\mathtt{Unknown}, \sum_{i\in\mathcal{M}}D_i=K, \mathtt{on}\right)$:}
After the indirect punishment is triggered, the attackers know their
own sensing results and will choose the actions as follows.
\begin{itemize}
\item \emph{Weak Cooperation ($\mathtt{Case.WC}$):} the attackers
will choose the action (N/A, wait) even if all attackers sense the
channel idle. Otherwise, they will receive a negative expected
aggregate reward $P_{M,K}^I-MP_{M,K}^BC_p$.
\item \emph{Strong Cooperation ($\mathtt{Case.SC}$):} the attackers
will choose the action (N/A, transmit) when all attackers sense the
channel idle. Even if at least one attackers senses the channel
busy, they will still choose the action (N/A, transmit) if
$P_{M,K}^I\geq MP_{M,K}^BC_p$.
\end{itemize}
\end{itemize}

\begin{lemma}\label{lemma:indirect_M_K}
The attackers' optimal long-term expected aggregate rewards in
$\mathtt{Case.WC}$ and $\mathtt{Case.SC}$ are
\begin{equation}
LR_{WC}^H=LR_{SC}^H=\frac{M}{1-\delta}Pr\left(\sum_{i\in\mathcal{N}}D_i=0\right)\left(P_{N,0}^I\frac{1}{N}-P_{N,0}^BC_p\right),
\end{equation}
by behaving honestly, and
\begin{equation}\label{eq:LR_WC_DH}
LR_{WC}^{DH}=\max_{0\leq z\leq
N}\frac{\sum_{k=0}^zPr(\sum_{i\in\mathcal{N}}D_i=k)(P_{N,k}^I-MP_{N,k}^BC_p)}{1-\delta(1-\sum_{k=0}^zPr(\sum_{i\in\mathcal{N}}D_i=k)P_{N,k}^B)}
\end{equation}
\begin{multline}\label{eq:LR_SC_DH}
LR_{SC}^{DH}=\max_{0\leq z\leq
N}{\sum_{k=0}^zPr(\sum_{i\in\mathcal{N}}D_i=k)}
[\frac{P_{N,k}^I-MP_{N,k}^BC_p}{1-\delta(1-\sum_{k=0}^zPr(\sum_{i\in\mathcal{N}}D_i=k)P_{N,k}^B)}
\\+\frac{\delta}{1-\delta}\frac{P_{N,k}^BPr(\sum_{i\in\mathcal{M}}D_i=0)(P_{M,0}^I-MP_{M,0}^BC_p)}{{1-\delta(1-\sum_{k=0}^zPr(\sum_{i\in\mathcal{N}}D_i=k)P_{N,k}^B)}}],
\end{multline}
by behaving dishonestly. Here the superscript ``$H$'' indicates
honest behaviors of attackers and ``$DH$'' indicates dishonest
behaviors. We denote the value of $z$ that achieves the maximum of
(\ref{eq:LR_WC_DH}) in $\mathtt{Case.WC}$ or (\ref{eq:LR_SC_DH}) in
$\mathtt{Case.SC}$ as $z^*$. The attackers' optimal policy $u^*$ has
a threshold structure as follows.
\begin{itemize}
\item \emph{If $\sum_{i\in\mathcal{N}}D_i\leq z^*$:} At least one attacker will
take the action (busy, transmit) before the indirect punishment is
triggered. After the indirect punishment is triggered,
\begin{itemize} \item $\mathtt{Case.WC}$: all attackers will take the action
(N/A, wait).
\item $\mathtt{Case.SC}$: all attackers will take the action (N/A, transmit)
if $P_{M,K}^I>MP_{M,K}^BC_p$ where $K$ attackers sense the channel
busy. Otherwise, they will take the action (N/A, wait).
\end{itemize}
\item \emph{If $\sum_{i\in\mathcal{N}}D_i> z^*$:} The attackers will
take the action (busy, wait) before the indirect punishment is
triggered. After the indirect punishment is triggered,
\begin{itemize} \item $\mathtt{Case.WC}$: all attackers will take the action
(N/A, wait).
\item $\mathtt{Case.SC}$: all attackers will take the action (N/A, transmit)
if $P_{M,K}^I>MP_{M,K}^BC_p$ where $K$ attackers sense the channel
busy. Otherwise, they will take the action (N/A, wait).
\end{itemize}
\end{itemize}
\end{lemma}

We can show in Lemma \ref{lemma:indirect_M_K} that
$LR_{WC}^{DH}<LR_{WC}^H$ in $\mathtt{Case.WC}$ and
$LR_{SC}^{DH}<LR_{SC}^H$ in $\mathtt{Case.SC}$ when $\delta$ is
close to $1$. Thus we can find an appropriate discount factor
threshold to ensure that $LR_{WC}^{DH}<LR_{WC}^H$ for
$\mathtt{Case.WC}$ and $LR_{SC}^{DH}<LR_{SC}^H$ for
$\mathtt{Case.SC}$.

\begin{theorem}
For multiple attackers in $\mathtt{Case.AT}$, there exists a
threshold $\delta^{th}\in(0,1)$ such that no attacks will happen if
$\delta>\delta^{th}$.
\end{theorem}

}

\subsection{Proof of Lemma \ref{lemma:longpunish1}}\label{subsec:Lemma_proof_indirect}
In $\mathtt{Case.NT}$, the attackers will only attack with the
action (busy, transmit) when all SUs sense the channel idle when the
indirect punishment is not triggered. But they may or may not
transmit after the punishment is triggered.

\subsubsection{$\mathtt{Case.WC}$ $($Weak Cooperation$)$}
In $\mathtt{Case.WC}$, the attackers will not transmit even when all
attackers sense the channel idle after the punishment is triggered.

If the attackers behave as honest SUs, the indirect punishment will
never be triggered. They share the spectrum opportunities with the
honest SUs when all SUs sense the channel idle. The attackers'
long-term expected aggregate reward is
$$
LR_{WC}^{H}=\sum_{t=0}^{\infty}\delta^tPr\left(\sum_{i\in\mathcal{N}}D_i=0\right)\left(P_{N,0}^I\frac{1}{N}-P_{N,0}^BC_p\right)M,
$$ which can be rewritten as in (\ref{eq:u_M_H}).

If the attackers attack with the action (busy, transmit) when all
SUs sense the channel idle ($\sum_{i\in\mathcal{N}}D_i=0$), then in
time slot $t=0$ the indirect punishment will be triggered with the
missed detection probability $P_{N,0}^B$. If no collision happens
with the probability $P_{N,0}^I$, the attack will not be detected
and the attackers will attack again if all SUs sense the channel
idle in the next time slot. By focusing on time slot $t=0$, the
attackers' long-term expected aggregate reward is
\begin{equation}
LR_{WC}^{DH}=Pr\left(\sum_{i\in\mathcal{N}}D_i>0\right)\delta
LR_{WC}^{DH} + Pr\left(\sum_{i\in\mathcal{N}}D_i=0\right) \big[
P_{N,0}^I\left(1+{\delta LR_{WC}^{DH}}\right)-P_{N,0}^B(MC_p) \big].
\nonumber
\end{equation}
We can then recursively rewrite $LR_{WC}^{DH}$  as in
(\ref{eq:u_M_NH}).

\subsubsection{$\mathtt{Case.SC}$ $($Strong Cooperation$)$}
In $\mathtt{Case.SC}$, the attackers will still transmit when all
attackers sense the channel idle after the punishment is triggered.

If the attackers behave as honest SUs, no indirect punishment will
be triggered and they will receive the same long-term expected
aggregate reward in (\ref{eq:u_M_H}).

If the attackers attack with the action (busy, transmit) when all
SUs sense the channel idle in time slot $t=0$, we can derive the
attackers' long-term expected aggregate reward similar as the
{$\mathtt{Case.WC}$},
\begin{align}
LR_{SC}^{DH} &=Pr\left(\sum_{i\in\mathcal{N}}D_i>0\right)\delta LR_{SC}^{DH}+Pr\left(\sum_{i\in\mathcal{N}}D_i=0\right) \nonumber \\
&\cdot \Big[ P_{N,0}^I(1+{\delta
LR_{SC}^{DH}})-P_{N,0}^BMC_p+P_{N,0}^B
\frac{\delta}{1-\delta}Pr\left(\sum_{i\in\mathcal{M}}D_i=0\right)(P_{M,0}^I-MP_{M,0}^BC_p))
\Big]. \nonumber
\end{align}
The only difference here is that the attackers can still obtain a
positive expected aggregate reward after the punishment is
triggered. Then we can recursively rewrite $LR_{SC}^{DH}$ as in
(\ref{eq:u_M_NH2}).



\end{document}